\documentclass[aps]{revtex4}

\usepackage{graphicx}

\def\calS{{\cal S}}

\begin{document}

\title{Nuclear Chemical and Mechanical Instability and 
the Liquid-Gas Phase Transition in Nuclei }

\author{ S.J. Lee$^{1,2}$ and A.Z. Mekjian$^1$}

\affiliation
{$^1$Department of Physics and Astronomy, Rutgers University,
Piscataway, NJ 08854}

\affiliation
{$^2$Department of Physics, Kyung Hee University, Yongin, KyungGiDo, Korea}

\begin{abstract}
The thermodynamic properties of nuclei are studied in a mean field model
using a Skryme interaction. 
Properties of two component systems are investigated
over the complete range of proton fraction from a system of pure neutrons to
a system of only protons. 
Besides volume, symmetry, and Coulomb effects we also
include momentum or velocity dependent forces. 
Applications of the results developed
are then given which include nuclear mechanical and chemical instability and
an associated liquid/gas phase transition in two component systems. 
The velocity dependence leads to further changes in the coexistence curve 
and nuclear mechanical and chemical instability curves.
\end{abstract}


\maketitle


\section{Introduction}

One primary goal of medium energy nuclear collisions is a detailed
study of the thermodynamic properties of strongly interacting
nuclear matter \cite{pr406,anp26}.
An important feature of these properties is the existence of
liquid-gas phase transition.
Properties of the nuclear force (long range attraction and
short range repulsion) parallel those of a van der Waals system \cite{prc27}
which qualitatively describes a liquid-gas phase transition
in atomic systems.
The liquid-gas phase transition in nuclei is the first phase
transition seen in a strongly interacting system.
Relativistic heavy ion collisions are being used to explore
a second phase transition from hadronic matter made of mesons
and baryons to a quark-gluon phase.
Important differences exist between the nuclear interaction
and interaction between atoms.
Because nuclei are made of neutrons and protons, the phase transition
is in a two component or binary system where symmetry energy
effects and Coulomb effects play an important role.
Moreover, the nuclear force has a velocity dependence.
The presence of symmetry energy and a Coulomb interaction effects 
and also a velocity dependence in the nuclear interaction makes
the nuclear case a unique and interesting binary system
within the general scope of such systems.
Examples of other two component systems are binary alloys and liquid $^3$He.
For $^3$He the two components are spin-up and spin down fluids.
The phase structure in such two component systems has some important
features. In nuclear systems, isospin fractionation 
\cite{pr406,anp26,isosp,prl85xu,prl85li,nucl2877}
is an example where the monomer gas phase has a large neutron
to proton ratio.
Ref.\cite{nucl2877} is the most recent reference to isospin fractionation
and contains further references to it.
Both the symmetry energy and Coulomb energy play an important role 
in this phenomena of isospin fractionation.
Nucleons carry spin but very little research has been done
in understanding the role of spin in the liquid gas phase structure.
However, the crust of neutron stars has features associated with
a superfluid phase.

An early study of the nuclear liquid-gas phase transition \cite{prc27}
treated the system as a one component system of nucleons.
This study was then extended to two components using a Skyrme 
interaction \cite{prc29}.
A relativistic mean field model was also developed in Ref.\cite{serot}
where the role of the symmetry energy was studied in detail.
The addition of the Coulomb energy \cite{prc63,plb580}
resulted in asymmetries which changed the mechanical and
chemical instability regions and binodal surface in pressure $P$,
temperature $T$ and proton fraction $y$ associated with phase coexistence.
For one component systems a phase diagram is the more familiar binodal
curve of pressure versus density or volume determined by a Maxwell construction.
Some other studies of one and two component phase transitions
can be found in 
Refs.\cite{prc65,prc67,prl89,npa748,epja25,prl95,sci298,gross,pr389,bertsch}.
The present work is an extension of our research reported 
in Refs.\cite{prc63,plb580}.
An extended Skyrme interaction is now used in our present
study and, for example, includes effects associated with a velocity
dependence in the nuclear interaction.
Here we will use a simplified form of the velocity dependence.
In particular we will use an effective mass approximation for it
which includes a density dependent behavior.
Our primary goal is to see what qualitative effects a momentum
dependence has when superimposed upon an interaction model
that does not include them.
A momentum dependence study was also given in Ref.\cite{plb650}
using a more refined dependence for it.
Our results differ from that of Ref.\cite{plb650}
since we also include Coulomb and surface effects.
Coulomb effects lead to an asymmetric behavior in proton fraction
\cite{prc63,plb580} of various quantities.
In the absence of Coulomb forces a symmetry exists around 
proton fraction $y=1/2$.
The velocity dependent force modifies nuclear saturation properties
and the symmetry energy.
Some recent extended studies of symmetry energy can be found 
in Refs.\cite{prc76,prc72}. 
The results given below show modifications in chemical and 
mechanical instability curves arising from an inclusion of
a density dependent effective mass.
The velocity dependence has a larger effect on the proton
rich instability and coexistence features compared to 
the neutron rich curves.
A detailed discussion is given below in Section \ref{rsltsec}
and in associated figures.
The study of two component nuclear systems with arbitrary neutron/proton 
ratios will be useful for future RIB (Rare Isotope Beam Facility) experiments 
and in astrophysical studies such as in neutron stars.

Our paper is divided as follows. The next section discusses the thermodynamic 
properties of nuclei. It is divided into two subsections. 
General results based on a mean field approach are presented in IIA. 
Specific results based on a Skyrme force for the potential terms, 
and low and high temperature kinetic energy behavior appear in IIB. 
This subsection also contains the effects of a velocity dependent interaction 
and related effective mass results. 
Then in Sect.\ref{rsltsec} we apply the results of 
subsection IIA and IIB to the specific issues of: A) mechanical and chemical 
instability of nuclei and B) the liquid-gas coexistence curve. 
Results are presented in 9 figures which are discussed. 
Finally, in Sect.IV a summary and conclusions are given.

\section{Themodynaic Properties of Nuclei in a Mean Field Description}

\subsection{General Results}

In this section we present results for the thermodynamic properties
of nuclear matter
which are extended from the results of Ref.\cite{prc63} to
include a velocity or momentum dependent interaction.
The matter is a two component system of protons and neutrons in
equilibrium at some temperature $T$. 
We first develop expressions for the total energy $E$ as a function
of the density $\rho_q$ of each component $q$ and temperature $T$.
The behavior of the energy functional with $\rho_q$ and $T$ can be used
to obtain the behavior of the pressure $P$ and chemical potential $\mu_q$
for each component of type $q$.
These quantities will also be functions of $\rho_q$ and $T$.
They can then be used to study, for example, a phase transition
in the nuclear system.

To begin, we use the fact that at a given temperature $T = 1/\beta$,
the proton and neutron constituents are distributed in phase space
according to the Wigner function $f$ as
\begin{eqnarray}
 f(\vec r, \vec p) = \sum_q f_q(\vec r, \vec p) ,   \hspace{1.5cm}
 f_q(\vec r, \vec p) = \frac{\gamma}{h^3} \tilde f_q(\vec r, \vec p)
    = \frac{\gamma}{h^3} \frac{1}{e^{\beta(\epsilon_q-\mu_q)} + 1}
            \label{wigf}
\end{eqnarray}
The spin degneracy factor $\gamma = 2$ and
$\epsilon_q$ and $\mu_q$ are the single particle energy
and the chemical potential of particle of type $q$.
Then the particle density $\rho$ and nucleon number $A$ are
given by the following equations:
\begin{eqnarray}
 \rho(\vec r) = \sum_q \rho_q(\vec r) , &\hspace{1.5cm}&
 \rho_q(\vec r) = \int d^3 p f_q(\vec r, \vec p)  ,   \\
 A = \sum_q N_q = \int d^3 r \rho(\vec r) , &\hspace{1.5cm}&
 N_q = \int d^3 r \rho_q(\vec r)
     = \int d^3 r \int d^3 p f_q(\vec r, \vec p)
\end{eqnarray}
Defining $\tau(\vec r)$ as 
\begin{eqnarray}
 \tau(\vec r) = \sum_q \tau_q(\vec r) , &\hspace{1.5cm}&
 \tau_q(\vec r) = \int d^3 p \frac{p^2}{\hbar^2} f_q(\vec r, \vec p) 
     \label{tauq}
\end{eqnarray}
the total energy $E$ is given by
\begin{eqnarray}
 E = \int d^3 r {\cal E}(\vec r)
   = \int d^3r \int d^3 p \frac{p^2}{2m} f(\vec r, \vec p)
    + \int d^3 r \int d^3 p U(\vec r, \vec p)
   = \int d^3 r \left[ {\cal E}_K(\vec r) + U(\vec r) \right]
           \label{tenerg}
\end{eqnarray}
The potential energy density is $U(\vec r)$, while
the ${\cal E}_K(\vec r) = \frac{\hbar^2}{2m}\tau(\vec r)$ 
is the kinetic energy density.
The single particle energy $\epsilon_q$ is given by
\begin{eqnarray}
 \epsilon_q &=& \frac{\delta E}{\delta f_q}
   = \frac{\delta {\cal E}(\vec r)}{\delta f_q(\vec r, \vec p)}
   = \frac{p^2}{2m} + \frac{\delta U}{\delta f_q}
   = \frac{p^2}{2m} + u_q(\vec r, \vec p)       \label{senerg} 
\end{eqnarray}
The $u_q = \frac{\delta U}{\delta f_q}$ is the single particle
potential of particle $q$ which may in general be momentum dependent.
The chemical potential $\mu_q$ is given by $\epsilon_q$ at an
effective Fermi momentum $p = p_{Fq}$ defined by the following equation:
\begin{eqnarray}
 \mu_q &=& \left.\epsilon_q\right|_{p=p_{Fq}}
   = \frac{p_{Fq}^2}{2m} + u_q(\vec r, \vec p_{Fq})  \label{chempot}
\end{eqnarray}

In order to study a phase transition we need information about the
behavior of the pressure when the system is in equilibrium.
The general expression for the pressure can be defined dynamically
from the total momentum conservation law,
$\frac{d}{dt} \left[\int d^3 r \int d^3 p ~\vec p f\right]
   = - \int d^3 r \vec\nabla_r \cdot \buildrel\leftrightarrow\over\Pi = 0$,
using the Vlasov equation as developed in Ref.\cite{bertsch}:
\begin{eqnarray}
 \frac{\partial f_q}{\partial t}
 + (\vec\nabla_p \epsilon_q) \cdot (\vec\nabla_r f_q)
 - (\vec\nabla_r \epsilon_q) \cdot (\vec\nabla_p f_q)
 = 0
\end{eqnarray}
A more general expression is obtained from the hydrodynamic consideration 
of TDHF in phase space as given in Ref.\cite{prc42q} which reads;
\begin{eqnarray}
\vec\nabla_r \cdot \buildrel\leftrightarrow\over\Pi
 &=& - \frac{d}{dt} \left[ \int d^3 p \vec p \sum_q f_q(\vec r, \vec p) \right]
  = - \sum_q \int d^3 p \vec p \left(\frac{\partial f_q}{\partial t}\right)
         \nonumber \\
 &=& \sum_q \int d^3 p \vec p
       ~\vec\nabla_r\cdot\left[(\vec\nabla_p\epsilon_q)f_q\right]
     + \sum_q \int d^3 p \hat{p} \cdot (\vec\nabla_r\epsilon_q) f_q
\end{eqnarray}
where $\hat p = \vec p / p$ is a unit vector in the direction of $\vec p$.
Using
$(\vec\nabla_r\epsilon_q)f_q = \vec\nabla_r(\epsilon_q f_q)
   - \epsilon_q \vec\nabla_r f_q
 = \vec\nabla_r(\epsilon_q f_q) - \vec\nabla_r {\cal E}$,
the dynamical pressure tensor $\Pi_{ij}$ is given by
\begin{eqnarray}
 \Pi_{ij} &=& \sum_q \int d^3 p p_i \left(\nabla_p^j \epsilon_q \right) f_q
 + \delta_{ij} \left[\int d^3 p \sum_q \epsilon_q f_q - {\cal E} \right]
           \nonumber \\
 &=& \sum_q \int d^3 p p_i
          \nabla_p^j \left(\frac{\delta {\cal E}}{\delta f_q}\right) f_q
 + \delta_{ij} \left[
   \sum_q \int d^3 p \left(\frac{\delta {\cal E}}{\delta f_q}\right) f_q
       - {\cal E} \right]   \nonumber \\
 &=& \sum_q \int d^3 p p_i \left[\frac{p_j}{m}
     + \nabla_p^j \left(\frac{\delta U}{\delta f_q}\right) \right] f_q
 + \delta_{ij} \left[\sum_q \int d^3 p
       \left(\frac{\delta U}{\delta f_q}\right) f_q
       - U \right]      \label{ppi}
\end{eqnarray}
Our previous study \cite{prc63} focused on 
a momentum independent potential which gave the following simpler
results for the pressure tensor:
\begin{eqnarray}
 \Pi_{ij} &=& \sum_q \int d^3 p \frac{p_i p_j}{m} f_q
    + \delta_{ij} \left[ \sum_q \left(\frac{\delta U}{\delta\rho_q}\right)
             \rho_q  - U \right]
    = \int d^3 p \sum_q \frac{p_i p_j}{m} f_q
    + \delta_{ij} \sum_q \rho \rho_q \frac{\delta(U/\rho)}{\delta\rho_q} 
\end{eqnarray}
The diagonal element of $\Pi_{ij}$ is the pressure $P = \Pi_{ij}$
which simplifies to
\begin{eqnarray}
 P = \Pi_{ii} = \sum_q \int d^3 p \frac{p_i^2}{m} f_q
     + \sum_q \frac{\delta U}{\delta\rho_q} \rho_q - U
   = P_K + \sum_q u_q\rho_q - U
   = P_K + P_P    \label{pres}
\end{eqnarray}
The $P_K = \int d^3 p \frac{p_i^2}{m} f = \frac{2}{3} {\cal E}_K$ is the
kinetic part of the pressure $P$, while the interaction potential part
is $P_P = \sum_q u_q \rho_q - U = \rho^2 \frac{\delta(U/\rho)}{\delta\rho}$. 
At temperature $T = 0$ the pressure $P$ is related to the derivative of
the energy per particle $E/A$ with particle number fixed as
\begin{eqnarray}
 P = \Pi_{ii} = - \frac{d (E/A)}{dV} = \rho^2 \frac{d({\cal E}/\rho)}{d\rho}
\end{eqnarray}
This result applies to a single component system.
Below we will give results at non zero temperature for a multi component system.
We first proceed with a discussion of the role of the momentum dependence
and effective mass.

As mentioned, our study is based on a qualitative study of the role
of a momentum dependent interaction and we therefore use a simplifying
approximation.
Specifically, we use an effective mass with a density dependence
and this approximation greatly simplifies our analysis in two
component asymmetric and finite nuclear systems.
We still include both Coulomb and surface effects since realistic
nuclear systems have such terms which are important in their
description and stability properties.
More refined stiudies will be developed in future work.
When the momentum dependent part is of the form 
 $A(\rho_p, \rho_n) \frac{p^2}{\hbar^2} f$, 
then it can be incorporated into the Hamiltonian as an effective mass term.
In Ref.\cite{plb650}, the momentum dependence is obtained from
\begin{eqnarray}
  \int \int d^3p d^3p' \frac{f_\tau(\vec r, \vec p) f_\tau'(\vec r, \vec p')}
         {1 + (\vec p - \vec p')^2/\Lambda^2}     \label{mompot}
\end{eqnarray}
We use an effective mass $m_q^*/m$ approach \cite{prc69} for Eq.(\ref{mompot}) 
which can further be approximated
by expanding the factor $1/(1 + (\vec p - \vec p')^2/\Lambda^2)$
to first order in $1 - (\vec p - \vec p')^2/\Lambda^2$.
Specifically, we write the effective mass behavior of $m/m^*_q$ as
\begin{eqnarray}
 \frac{m}{m_q^*} &=& 1 + A_q(\rho) \frac{2m}{\hbar^2}
            \hspace{2.0cm}
 A_q(\rho) \tau_q(\vec r) = \frac{\hbar^2}{2m_q^*} \tau_q(\vec r)
       - \frac{\hbar^2}{2m} \tau_q(\vec r)
\end{eqnarray}
Moreover, we have
 $U(\vec r) = U(\rho) + A(\rho) \tau(\vec r) 
  = U(\rho) + \sum_q A_q(\rho) \tau_q(\vec r)$
with $\tau_q(\vec r)$ 
of Eq.(\ref{tauq}).
A momentum dependent single particle potential $u_q(\vec r, \vec p)$ 
is given by
\begin{eqnarray}
 u_q(\vec r, \vec p) &=& \frac{\delta U(\vec r)}{\delta f_q(\vec r, \vec p)}
   = \frac{\delta U(\rho)}{\delta\rho_q} 
     + \frac{\delta A(\rho) \tau(\vec r)}{\delta\rho_q}  
     + A_q(\rho) \frac{p^2}{\hbar^2}   
   = \frac{\delta U(\vec r)}{\delta\rho_q}   
     + A_q(\rho) \frac{p^2}{\hbar^2}               
\end{eqnarray}
The $\mu_q$ is related to $u_q(\vec r, \vec p)$ through the result
\begin{eqnarray}
 \mu_q &=& \frac{p_{Fq}^2}{2m} + u_q(\vec r, \vec p_{Fq}) 
   = \left(1 + A_q(\rho) \frac{2m}{\hbar^2}\right) \frac{p_{Fq}^2}{2m}
     + \frac{\delta U(\rho)}{\delta\rho_q} 
     + \frac{\delta A(\rho) \tau(\vec r)}{\delta\rho_q}  
               \nonumber   \\
  &=& \left(1 + A_q(\rho) \frac{2m}{\hbar^2}\right) \frac{p_{Fq}^2}{2m}
      + \frac{\delta U(\vec r)}{\delta\rho_q}
   = \frac{p_{Fq}^2}{2m_q^*} + \frac{\delta U(\vec r)}{\delta\rho_q} 
               \label{muqem}   
\end{eqnarray}
Also
\begin{eqnarray}
\vec\nabla_p u_q(\vec r, \vec p) 
  &=& \vec\nabla_p \left(\frac{\delta U(\vec r)}{\delta f_q}\right)
   = A_q(\rho) \frac{2 \vec p}{\hbar^2}        
\end{eqnarray}
and
\begin{eqnarray}
 \int d^3 p u_q f_q &=& \int d^3 p 
       \left(\frac{\delta U(\vec r)}{\delta f_q}\right) f_q(\vec r, \vec p)
   = \frac{\delta U(\rho)}{\delta\rho_q} \rho_q(\vec r)
     + \frac{\delta A(\rho) \tau(\vec r)}{\delta\rho_q} \rho_q(\vec r) 
     + A_q(\rho) \tau_q(\vec r) 
               \nonumber   \\
  &=& \frac{\delta U(\vec r)}{\delta\rho_q(\vec r)} \rho_q(\vec r)
     + A_q(\rho) \tau_q(\vec r)
\end{eqnarray}
Here $\rho$ and $\tau$ are treated as independent variables. 
Then the pressure tensor $\Pi_{ij}$ is given by
\begin{eqnarray}
 \Pi_{ij} &=& \sum_q \int d^3 p \left(1 + A_q(\rho) \frac{2m}{\hbar^2}\right) \frac{p_i p_j}{m} f_q
    + \delta_{ij} \left[ \sum_q \left(\frac{\delta U(\vec r)}
           {\delta\rho_q}\right) \rho_q 
        + A(\rho) \tau(\vec r) - U(\vec r) \right]
                     \nonumber   \\
  &=& \int d^3 p \sum_q \left(1 + A_q(\rho) \frac{2m}{\hbar^2}\right) \frac{p_i p_j}{m} f_q
    + \delta_{ij} \sum_q \left[ 
      \rho(\vec r) \rho_q(\vec r) \frac{\delta(U(\rho)/\rho)} {\delta\rho_q}
    + \rho_q(\vec r) \frac{\delta A(\rho) \tau(\vec r)}{\delta\rho_q} \right]
                     \nonumber   \\
  &=& \int d^3 p \sum_q \frac{p_i p_j}{m_q^*} f_q
    + \delta_{ij} \sum_q \left[
       \rho(\vec r) \rho_q(\vec r) \frac{\delta(U(\vec r)/\rho)} {\delta\rho_q}
    + A(\rho) \tau(\vec r) \right]
\end{eqnarray}
and the pressure $P$ or diagonal element $\Pi_{ii} = P$ is
\begin{eqnarray}
 P &=& \Pi_{ii} 
    = \sum_q \left(1 + A_q(\rho) \frac{2m}{\hbar^2}\right) \int d^3 p \frac{p_i^2}{m} f_q
     + \sum_q \frac{\delta U(\vec r)}{\delta\rho_q} \rho_q 
         + A(\rho) \tau(\vec r) - U(\vec r)
               \nonumber   \\
   &=& \sum_q \int d^3 p \frac{p_i^2}{m_q^*} f_q
     + \sum_q \frac{\delta U(\vec r)}{\delta\rho_q} \rho_q
         - U(\rho)
   = P_K^* + P_P    \label{presem}
\end{eqnarray}
The $P_K^* = \sum_q P_{Kq}^*$ with
\begin{eqnarray}
 P_{Kq}^* &=& \left(1 + A_q(\rho) \frac{2m}{\hbar^2}\right) \int d^3 p \frac{p_i^2}{m} f_q 
   = \int d^3 p \frac{p^2}{3 m_q^*} f_q 
   = \frac{2}{3} {\cal E}_K^*
\end{eqnarray}
is the kinetic pressure with an effective mass correction term,
and the second equality is for an isotropic momentun distribution.
The potential part of the pressure $P_P$ is given by 
\begin{eqnarray} 
 P_P &=& \sum_q \frac{\delta U(\vec r)}{\delta\rho_q} \rho_q - U(\rho) 
  = \sum_q \left[\rho \rho_q \frac{\delta (U(\rho)/\rho)}{\delta\rho_q}  
   + \rho_q \frac{\delta A(\rho) \tau(\vec r)}{\delta\rho_q}\right]  
  = \sum_q \left[\rho \rho_q \frac{\delta (U(\vec r)/\rho)}{\delta\rho_q}
   + A_q(\rho) \tau_q(\vec r)\right]
              \nonumber   \\
 &=& \rho^2 \frac{\delta (U(\rho)/\rho)}{\delta\rho}
   + \rho \frac{\delta A(\rho) \tau(\vec r)}{\delta\rho}
  = \rho^2 \frac{\delta (U(\vec r)/\rho)}{\delta\rho}
   + A(\rho) \tau(\vec r)
  = \rho^2 \frac{\delta (U(\vec r)/\rho)}{\delta\rho}
   + {\cal E}_K^* - {\cal E}_K
\end{eqnarray}
The ${\cal E}_K = \sum_q \frac{\hbar^2}{2m} \tau_q(\vec r)$ and 
\begin{eqnarray}
 {\cal E}_K^* &=& \sum_q \frac{\hbar^2}{2m_q^*} \tau_q(\vec r) 
    = \sum_q \left(1 + A_q(\rho) \frac{2m}{\hbar^2}\right) 
      \frac{\hbar^2}{2m} \tau_q(\vec r) .
\end{eqnarray}
Also in obtaining this result we use the fact that
\begin{eqnarray}
 \sum_q \rho_q \frac{\delta U(\rho, \rho_q)}{\delta\rho_q}
  &=& \rho \frac{\delta U(\rho, \rho x_q)}{\delta\rho}
\end{eqnarray}
which can be shown by looking at the derivative
of $B(\rho) C(\rho_p) D(\rho_n) = B(\rho) C(\rho x_p) D(\rho x_n)$.
Here the variation $\rho$ must be done after 
replacing $\rho_q$ by $\rho x_q$.

Other thermodynamic variables, such as $S$, $\Omega$, $F$, $G$ are
given in Ref.\cite{prc63}.
The entropy $S$ follows, from the distribution $\tilde f_q$ of Eq.(\ref{wigf}),
\begin{eqnarray}
 S &=& \sum_q S_q = \int d^3 r \calS = \int d^3 r \sum_q \calS_q 
\end{eqnarray}
and
\begin{eqnarray}
 \calS_q &=& - \frac{\gamma}{h^3} \int d^3 p
     \left[\tilde f_q \ln \tilde f_q + (1-\tilde f_q) \ln (1-\tilde f_q)\right]
                \nonumber \\
   &=& \beta \int d^3 p \epsilon_q f_q
   + \beta \int d^3 p
         \frac{\vec p\cdot\vec\nabla_p\epsilon_q}{3} f_q
   - \beta \mu_q \int d^3 p f_q     \label{sqden}
\end{eqnarray}
In equilibrium, from Eqs.(\ref{ppi}) and (\ref{sqden})
\begin{eqnarray}
 T \calS 
  &=& {\cal E} + P - \sum_q \mu_q \rho_q  
       = {\cal E}_K + P_K - \sum_q (\mu_q - u_q) \rho_q  
              \nonumber   \\
  &=& {\cal E}_K + P_K - \sum_q \frac{p_{Fq}^2}{2m} \rho_q  ,  \label{ts0}  
\end{eqnarray}
The last eqaulity of Eq.(\ref{ts0}) is the result of using Eq.(\ref{chempot}).
For momentum dependent potential the entropy is now
\begin{eqnarray}
 T \calS &=& {\cal E} + P - \sum_q \mu_q \rho_q  
   = {\cal E}_K^* + P_K^* 
     - \sum_q \left(\mu_q - \frac{\delta U(\vec r)}{\delta\rho_q}\right) \rho_q 
            \nonumber   \\
  &=& {\cal E}_K^* + P_K^*
     - \sum_q \left(1 + A_q(\rho) \frac{2m}{\hbar^2}\right) \frac{p_{Fq}^2}{2m} \rho_q 
   = {\cal E}_K^* + P_K^* - \sum_q \frac{p_{Fq}^2}{2m_q^*} \rho_q
              \label{tsem}  
\end{eqnarray}
where use has been made of Eqs.(\ref{muqem}) and (\ref{presem}) 
to obtain this result.
General thermodynamic relations also determine the entropy, 
pressure and chemical potential \cite{prc63}.

\subsection{Thermodynamic Properties of Nuclear Matter based on a Skyrme
Interaction}

We now use a Skyrme interaction to develop expressions for the potential $U$.
Once the potential energy $U$ in Eq.(\ref{tenerg}) is known,
then questions related to mechanical and chemical instability 
and the possibility of a phase transition of the system
can be studied  using Eqs.(\ref{wigf}) -- (\ref{tsem}).  
The potential energy $U$ determines $\epsilon_q$ and $\mu_q$ and
the potential energy part of $E$ and $P$.
Then for fixed $T$ and $N_q$, the Wigner function $f$ and $p_{Fq}$ are
determined and thus the kinetic terms of $E$, $\mu_q$, and $P$.
Using these results, the entropy $\calS$ can be determined.
For a nuclear system of proton ($\rho_p$) and neutron ($\rho_n$),
this gives the local potential energy density as
\begin{eqnarray}
 U(\rho_q) &=& \frac{t_0}{2} \left(1 + \frac{x_0}{2}\right) \rho^2
         - \frac{t_0}{2} \left(\frac{1}{2} + x_0\right) \sum_q \rho_q^2
     + \frac{t_3}{12} \left(1 + \frac{x_3}{2}\right) \rho^{\alpha+2}
         - \frac{t_3}{12} \left(\frac{1}{2} + x_3\right) 
            \rho^\alpha \sum_q \rho_q^2
              \nonumber   \\
  & & + \frac{1}{4} \left[t_1 \left(1 + \frac{x_1}{2}\right)
         + t_2 \left(1 + \frac{x_2}{2}\right)\right] \rho \tau
      - \frac{1}{4} \left[t_1 \left(\frac{1}{2} + x_1\right)
         - t_2 \left(\frac{1}{2} + x_2\right)\right] \sum_q \rho_q \tau_q
              \nonumber   \\  & &
         + C \rho^\beta \rho_p^2 + C_s \rho^\eta
    \label{potene}
\end{eqnarray}
Here $C \rho^\beta = \frac{4\pi}{5} e^2 R^2$
and $C_s \rho^\eta = \frac{4\pi R^2 \sigma(\rho)}{V}
  = \frac{(4\pi r_0^2 \sigma)}{V^{1/3}} \rho^{2/3}$
when we approximate the Coulomb and surface effects as
coming from a finite uniform sphere of radius $R = r_0 A^{1/3}$
with total charge $Z$ ($U_C = \frac{3}{5}\frac{e^2 Z^2}{R V}$) \cite{prc63}.
The values for the force parameters used here are given in Table \ref{tabl1}.
\begin{table}
\caption{Skyrme parameters used here are in MeV and fm units \cite{plb580}.
For $t_1$ and $t_2$, the SkM parameter values are used.
  }
  \label{tabl1}
\begin{tabular}{cccccc}
\hline
 $t_0$ \ \ \ \ \ \ \ \ $x_0$ & $t_3$ \ \ \ \ \ \ \ \ $x_3$ &  $\alpha$   \\
 --1089.0 \ \ \ --1/6 \ \ \ &   17480.4 \ \ \ --1/2 \ \ \  &   1   \\
\hline
            & \ \ Momentum-Dep. \ \  & \ \ Momentum-Indep. \ \ \\
   $t_1$    &    251.11           &     0     \\
   $x_1$    &    --1/2            &    --1/2  \\
   $t_2$    &   --150.66          &     0     \\
   $x_2$    &    --1/2            &    --1/2  \\
 Effective mass $m^*/m$  &    0.895626         &     1     \\
 Binding energy $E_B/A$  &    13.1057          &  15.54447 \\
 Fermi energy   $E_F$    &    31.8018          &  34.2101  \\
 Saturation density $\rho_0$ &    0.1283           &  0.143145 \\
 Symmetry energy $S_V$   &    23.4791          &  24.39379 \\
 Compresibility $\kappa$ &    307.780          &  361.9045 \\
\hline
\end{tabular}
\end{table}
We define an effective mass $m_q^*$ as
\begin{eqnarray}
 \frac{m}{m_q^*} &=& 1 + \frac{2m}{\hbar^2} \left\{
        \frac{1}{4} \left[t_1 \left(1 + \frac{x_1}{2}\right)
         + t_2 \left(1 + \frac{x_2}{2}\right)\right] \rho 
      - \frac{1}{4} \left[t_1 \left(\frac{1}{2} + x_1\right)
         - t_2 \left(\frac{1}{2} + x_2\right)\right] \rho_q \right\}
\end{eqnarray} 
Then the momentum dependent potential term becomes
\begin{eqnarray}
 A_q(\rho) &=& \frac{\hbar^2}{2m_q^*} - \frac{\hbar^2}{2m}
   = \frac{1}{4} \left[t_1 \left(1 + \frac{x_1}{2}\right)
       + t_2 \left(1 + \frac{x_2}{2}\right)\right] \rho 
      - \frac{1}{4} \left[t_1 \left(\frac{1}{2} + x_1\right)
       - t_2 \left(\frac{1}{2} + x_2\right)\right] \rho_q 
              \nonumber   \\  
  &=& \frac{\hbar^2}{2m} \left[ -1 + 1 + \frac{2m}{\hbar^2} \left\{
        \frac{1}{4} \left[t_1 \left(1 + \frac{x_1}{2}\right)
       + t_2 \left(1 + \frac{x_2}{2}\right)\right] \rho 
      - \frac{1}{4} \left[t_1 \left(\frac{1}{2} + x_1\right)
       - t_2 \left(\frac{1}{2} + x_2\right)\right] \rho_q \right\} \right]
\end{eqnarray}
For a symmetric nucleus, $N = Z$, $\rho_q = \rho/2$,
and thus
\begin{eqnarray}
 U(\rho) = \frac{3}{8} t_0 \rho^2 + \frac{3}{48} t_3 \rho^{\alpha+2}
         + \frac{3}{16} \left(t_1 + t_2\right) \rho \tau
         + C \rho^\beta \rho_p^2 + C_s \rho^\eta
\end{eqnarray}
This potential enegy determines the interaction dependent terms
of ${\cal E}$, $P$, $\epsilon_q$, and $\mu_q$
which depend on densities without an explicit $T$ dependence.

For a momentum dependent potential energy as in Eq.(\ref{potene}),
$\epsilon_q - \mu_q = (p^2 - p_{Fq}^2)/(2m_q^*)$ 
where the effective mass $m_q^*$
is independent of the momentum independent part of potential and
the Wigner funcion of Eq.(\ref{wigf}) becomes
\begin{eqnarray}
 \tilde f_q(\vec r, \vec p) = \frac{1}{e^{\beta(\epsilon_q - \mu_q)} + 1}
       = \frac{1}{e^{\beta(p^2 - p_{Fq}^2)/(2m_q^*)} + 1}
\end{eqnarray}
Thus we can evaluate the kinetic terms in ${\cal E}$,
$P$, and $\mu_q$ which are functions of $T$ and $p_{Fq}$.
Defining the Fermi integral $F_\alpha(\eta)$, with effective mass $m_q^*$,
\begin{eqnarray}
 F_\alpha(\eta_q) &=& \int_0^\infty \frac{x^\alpha}{1 + e^{x-\eta_q}} dx
     = \left(\frac{\lambda_q^2}{4\pi\hbar^2}\right)^{\alpha+1}
       \int_0^\infty \frac{2 p^{2\alpha+1} dp}
             {1 + e^{\beta p^2/{2m_q^*} - \eta_q}}  ,   \\
 \eta_q &=& \beta \left(\mu_q - \frac{\delta U(\vec r)}{\delta\rho_q}\right)
       = \beta p_{Fq}^2/(2m_q^*) = p_{Fq}^2/(2m_q^*T) = \ln z_q   ,   \\
 \lambda_q &=& \sqrt{2\pi\hbar^2/m_q^*T}
\end{eqnarray}
we can write, for $f(\vec r, \vec p) = f(\vec r, p)$,
\begin{eqnarray}
 \rho_q &=& \int d^3 p f_q(\vec r, \vec p)
   = \frac{\gamma}{h^3} \int d^3 p \frac{1}{e^{\beta(p^2-p_{Fq}^2)/(2m_q^*)} + 1}
   = \lambda_q^{-3} \frac{2\gamma}{\sqrt{\pi}} F_{1/2}(\eta_q) , \label{rhoq} \\
 \epsilon_{Fq}^* &\equiv& \frac{p_{Fq}^2}{2m_q^*}
   = \frac{\hbar^2}{2m_q^*}
         \left(\frac{6\pi^2}{\gamma}\rho_q\right)^{2/3} 
   = \frac{m}{m_q^*} \epsilon_{Fq}  ,   
          \hspace{2.0cm}
 \epsilon_{Fq} = \frac{\hbar^2}{2m} 
      \left(\frac{6\pi^2}{\gamma}\rho_q\right)^{2/3} ,      \\
 \tau_q &=& \int d^3 p \frac{p^2}{\hbar^2} f_q(\vec r, \vec p)
  = \frac{\gamma}{h^3} \int d^3 p \frac{p^2}{\hbar^2}
        \frac{1}{e^{\beta(p^2-p_{Fq}^2)/(2m_q^*)} + 1}     \nonumber \\
  &=& 8\gamma \sqrt{\pi} \lambda_q^{-5} F_{3/2}(\eta_q)
   = \frac{1}{\beta} \frac{2m_q^*}{\hbar^2} \frac{2\gamma}{\sqrt{\pi}}
                 \lambda_q^{-3} F_{3/2}(\eta_q)  
   = \frac{2m}{\hbar^2} {\cal E}_{Kq}      
   = \frac{2m_q^*}{\hbar^2} {\cal E}_{Kq}^* ,   \\  
 {\cal E}_{Kq} &=& \frac{\hbar^2}{2m} \tau_q = \frac{3}{2} P_{Kq}
  = \int d^3 p \frac{p^2}{2m} f_q(\vec r, \vec p)
  = \frac{\gamma}{h^3} \int d^3 p \frac{p^2}{2m}
        \frac{1}{e^{\beta(p^2-p_{Fq}^2)/(2m_q^*)} + 1}     \nonumber \\
  &=& \frac{4\gamma\hbar^2\sqrt{\pi}}{m}
                 \lambda_q^{-5} F_{3/2}(\eta_q)
   = \frac{m_q^*}{m} \frac{1}{\beta} \frac{2\gamma}{\sqrt{\pi}}
                 \lambda_q^{-3} F_{3/2}(\eta_q) ,    \\  
 {\cal E}_{Kq}^* &=& \frac{\hbar^2}{2m_q^*} \tau_q
  = \frac{3}{2} P_{Kq}^*
  = \int d^3 p \frac{p^2}{2m_q^*} f_q(\vec r, \vec p)
  = \frac{\gamma}{h^3} \int d^3 p \frac{p^2}{2m_q^*}
        \frac{1}{e^{\beta(p^2-p_{Fq}^2)/(2m_q^*)} + 1}     \nonumber \\
  &=& \frac{4\gamma\hbar^2\sqrt{\pi}}{m_q^*}
                 \lambda_q^{-5} F_{3/2}(\eta_q)
   = \frac{1}{\beta} \frac{2\gamma}{\sqrt{\pi}}
                 \lambda_q^{-3} F_{3/2}(\eta_q)      
\end{eqnarray}
Here $\epsilon_{Fq}$ is the chemical potential at absolute zero 
or Fermi energy and $p_{Fq}$ is the effective Fermi momentum at $T$ 
(which is related to density $\rho_q$ through Eq.(\ref{rhoq})).
The particle number $N_q = \int d^3 r \rho(\vec r)$ determines
the effective Fermi momentum $p_{Fq}(\vec r)$ or $\eta_q$ at $T$,
in terms of density $\rho_q(\vec r)$,
\begin{eqnarray}
 \eta_q(\rho_q, T) 
   &=& \beta \left(\mu_q - \frac{\delta U(\vec r)}{\delta\rho_q}\right) 
     = \beta\frac{p_{Fq}^2}{2m_q^*}
     = F_{1/2}^{-1}\left(\frac{\sqrt{\pi}}{2\gamma} \lambda_q^3 \rho_q\right)
\end{eqnarray}

For multi(two)-component systems with potential energy given
by Eq.(\ref{potene}), with a given $\rho_q$ (or $p_{Fq}$) and $T$,
the thermodynamic properties are as follows.
The chemical potential is given by
\begin{eqnarray}
 \mu_q(\rho_q,T) &=& T \eta_q(\rho_q, T) 
        + \frac{\delta U(\vec r)}{\delta\rho_q}
         \nonumber \\  
  &=& T \eta_q(\rho_q, T) 
   + t_0 \left(1+\frac{x_0}{2}\right) \rho
      + \frac{t_3}{12} \left(1+\frac{x_3}{2}\right) (\alpha+2) \rho^{\alpha+1}
   - \frac{t_3}{12} \left(\frac{1}{2}+x_3\right) \alpha \rho^{\alpha+1}
       \nonumber \\  & &
   - t_0 \left(\frac{1}{2}+x_0\right) \rho_q
 + \frac{t_3}{12} \left(\frac{1}{2}+x_3\right) (\alpha-1) 2 \rho^\alpha \rho_q
 - \frac{t_3}{12} \left(\frac{1}{2}+x_3\right) 2\alpha \rho^{\alpha-1} \rho_q^2
         \nonumber \\  & &
   + \frac{1}{4} \left[t_1 \left(1 + \frac{x_1}{2}\right)
        + t_2 \left(1 + \frac{x_2}{2}\right)\right] \tau
   - \frac{1}{4} \left[t_1 \left(\frac{1}{2} + x_1\right)
        - t_2 \left(\frac{1}{2} + x_2\right)\right] \tau_q
              \nonumber   \\  & &
  + C \beta\rho^{\beta-1} \rho_p^2 + 2 C \rho^\beta \rho_p \delta_{q,p}
               + \eta C_s \rho^{\eta-1}   . 
\end{eqnarray}
The equation of state has a behavior determined by
\begin{eqnarray}
 P(\rho_q,T) &=& \sum_q \frac{2}{3} {\cal E}_{Kq}^*(\rho_q,T)  
         + \rho^2 \frac{\delta(U(\rho)/\rho)}{\delta\rho}
         + \rho \frac{\delta A(\rho)\tau(\vec r)}{\delta\rho} 
                 \nonumber \\
  &=& \sum_q \left[\frac{5}{3} {\cal E}_{Kq}^*(\rho_q,T) 
              - {\cal E}_{Kq}(\rho_q,T)\right]
         + \rho^2 \frac{\delta(U(\vec r)/\rho)}{\delta\rho}
                 \nonumber \\
  &=& \sum_q \left[\frac{5}{3} {\cal E}_{Kq}^*(\rho_q,T)  
                  - {\cal E}_{Kq}(\rho_q,T)\right]
   + \frac{t_0}{2} \left(1 + \frac{x_0}{2}\right) \rho^2
   + \frac{t_3}{12} \left(1 + \frac{x_3}{2}\right)(\alpha+1) \rho^{\alpha+2}
       \nonumber \\  & &
   - \frac{t_0}{2} \left(\frac{1}{2} + x_0\right) \sum_q \rho_q^2
   - \frac{t_3}{12} \left(\frac{1}{2} + x_3\right)
        (\alpha+1) \rho^\alpha \sum_q \rho_q^2
         \nonumber \\  & &
   + C (\beta + 1) \rho^\beta \rho_p^2  + C_s (\eta - 1) \rho^\eta . 
\end{eqnarray}
The energy density is
\begin{eqnarray}
 {\cal E}(\rho_q,T) &=& \sum_q {\cal E}_{Kq}(\rho_q,T) + U(\vec r)    
    = \sum_q {\cal E}_{Kq}^*(\rho_q,T) + U(\rho)    \nonumber   \\
   &=& \sum_q {\cal E}_{Kq}^*(\rho_q,T)
       \nonumber \\  & &
    + \frac{t_0}{2} \left(1 + \frac{x_0}{2}\right) \rho^2
              - \frac{t_0}{2} \left(\frac{1}{2} + x_0\right) \sum_q \rho_q^2
         + \frac{t_3}{12} \left(1 + \frac{x_3}{2}\right) \rho^{\alpha+2}
   - \frac{t_3}{12} \left(\frac{1}{2} + x_3\right) \rho^\alpha \sum_q \rho_q^2
                 \nonumber \\  & &
      + C \rho^\beta \rho_p^2 + C_s \rho^\eta   
\end{eqnarray}
and the entropy can be obtained from
\begin{eqnarray}
 T\calS(\rho_q,T) &=& \sum_q \frac{5}{3} {\cal E}_{Kq}^*(\rho_q,T)
                - \sum_q (\mu_q - \frac{\delta U(\vec r)}{\delta\rho_q}) \rho_q 
     = \sum_q \frac{5}{3} {\cal E}_{Kq}^*(\rho_q,T)
        - T \sum_q \eta_q(\rho_q, T) \rho_q . 
\end{eqnarray}
Once we evaluate $F_{1/2}(\eta)$ and $F_{3/2}(\eta)$,
or more directly $\eta = F_{1/2}^{-1}(\chi)$ and $F_{3/2}(\eta)$,
we can evaluate various thermodynamic quantities
in terms of $\rho_q$ and $T$.

For low temperature and high density limit, $\lambda^3\rho$ large,
i.e., when the average de Broglie thermal wavelength $\lambda$ is larger
than the average interparticle separation $\rho^{-1/3}$,
we can use a nearly degenerate (Fermi gas) approximations \cite{huang}
for $F_{1/2}$ to obtain
\begin{eqnarray}
 \eta_q(\rho_q, T) 
  &=&  \beta \left(\mu_q - \frac{\delta U(\vec r)}{\delta\rho_q}\right) 
     = \beta\frac{p_{Fq}^2}{2m_q^*}
     = F_{1/2}^{-1}\left(\frac{\sqrt{\pi}}{2\gamma} \lambda_q^3 \rho_q\right)
     = \beta \epsilon_{Fq}^* \left[ 1
             - \frac{\pi^2}{12} \left(\frac{T}{\epsilon_{Fq}^*} \right)^2
             + \cdots \right]      \nonumber  \\
    &=& \beta \frac{\hbar^2}{2m_q^*}\left(\frac{6\pi^2}{\gamma}\right)^{2/3}
      \left[ \rho_q^{2/3}
       - \frac{\pi^2 {m_q^*}^2}{3\hbar^4} \left(\frac{\gamma}{6\pi^2}\right)^{4/3}
               T^2 \rho_q^{-2/3} + \cdots \right]   ,   \\
 {\cal E}_{Kq}^*(\rho_q, T) &=& \frac{2\gamma}{\beta\sqrt{\pi}}
                 \lambda_q^{-3} F_{3/2}(\eta_q)
    = \frac{3}{2} P_{Kq}^*
    = \frac{3}{5} \rho_q\epsilon_{Fq}^* \left[ 1
         + \frac{5\pi^2}{12} \left(\frac{T}{\epsilon_{Fq}^*}\right)^2
             + \cdots \right]      \nonumber  \\
   &=& \frac{3\hbar^2}{10m_q^*}\left(\frac{6\pi^2}{\gamma}\right)^{2/3}
      \left[ \rho_q^{5/3} + \frac{5\pi^2 {m_q^*}^2}{3\hbar^4}
               \left(\frac{\gamma}{6\pi^2}\right)^{4/3}
               T^2 \rho_q^{1/3} + \cdots \right]   ,   \\
 \tau_q(\rho_q, T) &=& \frac{2m_q^*}{\hbar^2} {\cal E}_{Kq}^*
   = \frac{3}{5} \frac{2m_q^*}{\hbar^2} \rho_q\epsilon_{Fq}^* \left[ 1
         + \frac{5\pi^2}{12} \left(\frac{T}{\epsilon_{Fq}^*}\right)^2
             + \cdots \right]      \nonumber \\
  &=& \frac{3}{5} \left(\frac{6\pi^2}{\gamma}\right)^{2/3} \rho_q^{5/3} \left[1
         + \frac{5\pi^2}{12} \left(\frac{T}{\epsilon_{Fq}^*}\right)^2
             + \cdots \right]      \nonumber  \\
  &=& \frac{3}{5}\left(\frac{6\pi^2}{\gamma}\right)^{2/3}
      \left[ \rho_q^{5/3} + \frac{5\pi^2 {m_q^*}^2}{3\hbar^4}
               \left(\frac{\gamma}{6\pi^2}\right)^{4/3}
               T^2 \rho_q^{1/3} + \cdots \right]   
\end{eqnarray}
In the other limit where $\lambda_q^3\rho$ is small,
we have a nearly non-degenerate Fermi gas (classical ideal gas)
and the resulting equations are given by an ideal gas
in leading order with higher order corrections \cite{huang} as
\begin{eqnarray}
 \eta_q(\rho_q, T) 
  &=& \beta \left(\mu_q - \frac{\delta U(\vec r)}{\delta\rho_q}\right)
    = \ln\left[\frac{\rho_q\lambda_q^3}{\gamma}
       \left(1 + \frac{1}{2\sqrt{2}} \frac{\rho_q\lambda_q^3}{\gamma}
         + \cdots \right) \right]
   \approx \ln\left(\frac{\rho_q\lambda_q^3}{\gamma}\right)
       + \frac{1}{2\sqrt{2}} \left(\frac{\rho_q\lambda_q^3}{\gamma}\right)
                 ,    \label{etaqht}   \\
 {\cal E}_{Kq}^*(\rho_q, T) &=& \frac{3}{2} P_{Kq}^*
    = \frac{3}{2} \rho_q T\left[1
      + \frac{1}{2^{5/2}} \frac{\rho_q\lambda_q^3}{\gamma}
      + \left(\frac{1}{8} - \frac{2}{3^{5/2}}\right)
             \left(\frac{\rho_q\lambda_q^3}{\gamma}\right)^2
      + \cdots \right]    \label{enqht}   ,   \\
 \tau_q(\rho_q, T) &=& \frac{2m_q^*}{\hbar^2} {\cal E}_{Kq}^*
    = \frac{2m_q*}{\hbar^2} \frac{3}{2} \rho_q T\left[1
      + \frac{1}{2^{5/2}} \frac{\rho_q\lambda_q^3}{\gamma}
      + \left(\frac{1}{8} - \frac{2}{3^{5/2}}\right)
             \left(\frac{\rho_q\lambda_q^3}{\gamma}\right)^2
      + \cdots \right]    \label{tauht}
\end{eqnarray}

For a nuclear system with protons and neutrons with the interaction
given by Eq.(\ref{potene}),
the non-degenerate Fermi gas limit of Eqs.(\ref{etaqht}), (\ref{enqht})
and (\ref{tauht}) leads to the following set of equations.
The chemical potential has a behavior determined by
\begin{eqnarray}
 \mu_q(\rho,y,T) &=& T \ln\left[ \left(\frac{\lambda_q^3}{\gamma}\right)
                               \rho_q \right]
    + \frac{T}{2\sqrt{2}} \left(\frac{\lambda_q^3}{\gamma}\right) \rho_q
           \nonumber \\  & &
   + \frac{1}{4} \left[t_1 \left(1 + \frac{x_1}{2}\right)
       + t_2 \left(1 + \frac{x_2}{2}\right)\right] 
          \frac{3}{2} T \sum_q \frac{2m_q^*}{\hbar^2} 
          \left[\rho_q + \frac{\lambda_q^3}{2^{5/2} \gamma} \rho_q^2\right]
              \nonumber   \\  & &
   - \frac{1}{4} \left[t_1 \left(\frac{1}{2} + x_1\right)
       - t_2 \left(\frac{1}{2} + x_2\right)\right] 
          \frac{3}{2} T \frac{2m_q^*}{\hbar^2} 
          \left[\rho_q + \frac{\lambda_q^3}{2^{5/2} \gamma} \rho_q^2\right]
              \nonumber   \\  & &
   + t_0 \left(1+\frac{x_0}{2}\right) \rho
      + \frac{t_3}{12} \left(1+\frac{x_3}{2}\right) (\alpha+2) \rho^{\alpha+1}
   - \frac{t_3}{12} \left(\frac{1}{2}+x_3\right) \alpha \rho^{\alpha+1}
       \nonumber \\  & &
   - t_0 \left(\frac{1}{2}+x_0\right) \rho_q
 + \frac{t_3}{12} \left(\frac{1}{2}+x_3\right) (\alpha-1) 2 \rho^\alpha \rho_q
 - \frac{t_3}{12} \left(\frac{1}{2}+x_3\right) 2\alpha \rho^{\alpha-1} \rho_q^2
         \nonumber \\  & &
  + C \beta\rho^{\beta-1} \rho_p^2 + 2 C \rho^\beta \rho_p \delta_{q,p}
    + \eta C_s \rho^{\eta-1} .  \label{murqt} 
\end{eqnarray}
The equation of state has a form given by
\begin{eqnarray}
 P(\rho,y,T) &=& \frac{5}{2} T\rho
   + \frac{5}{2} \frac{T}{2\sqrt{2}} \sum_q 
       \left(\frac{\lambda_q^3}{\gamma}\right) \left(\frac{\rho_q^2}{2}\right)
   - \frac{3}{2} T \sum_q \frac{m_q^*}{m} \left[\rho_q
        + \frac{1}{2\sqrt{2}} \left(\frac{\lambda_q^3}{\gamma} \right)
            \left(\frac{\rho_q^2}{2}\right)\right]
                     \nonumber \\  & &
   + \frac{t_0}{2} \left(1 + \frac{x_0}{2}\right) \rho^2
   + \frac{t_3}{12} \left(1 + \frac{x_3}{2}\right)(\alpha+1) \rho^{\alpha+2}
       \nonumber \\  & &
   - \frac{t_0}{2} \left(\frac{1}{2} + x_0\right) \sum_q \rho_q^2
   - \frac{t_3}{12} \left(\frac{1}{2} + x_3\right) (\alpha+1) 
         \rho^\alpha \sum_q \rho_q^2
         \nonumber \\  & &
   + C (\beta + 1) \rho^\beta \rho_p^2  + C_s (\eta - 1) \rho^\eta
               \label{prqt} .  
\end{eqnarray}
The energy density is
\begin{eqnarray}
 {\cal E}(\rho,y,T) &=& \frac{3}{2} T \rho
   + \frac{3}{2} \frac{T}{2\sqrt{2}} \sum_q
     \left(\frac{\lambda_q^3}{\gamma}\right) \left(\frac{\rho_q^2}{2}\right)
                     \nonumber \\  & &
   + \frac{t_0}{2} \left(1 + \frac{x_0}{2}\right) \rho^2
   - \frac{t_0}{2} \left(\frac{1}{2} + x_0\right) \sum_q \rho_q^2
   + \frac{t_3}{12} \left(1 + \frac{x_3}{2}\right) \rho^{\alpha+2}
   - \frac{t_3}{12} \left(\frac{1}{2} + x_3\right) \rho^\alpha \sum_q \rho_q^2
                 \nonumber \\  & &
   + C \rho^\beta \rho_p^2 + C_s \rho^\eta
               \label{erqt}   
\end{eqnarray}
and the entropy is
\begin{eqnarray}
 T\calS(\rho,y,T) &=& \frac{5}{2} T \rho
        - T \sum_q \rho_q \ln\left(\frac{\lambda_q^3}{\gamma} \rho_q\right)
      + \frac{T}{2\sqrt{2}} \sum_q 
        \left(\frac{\lambda_q^3}{\gamma}\right) \left(\frac{\rho_q^2}{4}\right)
               \label{tsrqt}
\end{eqnarray}
The effective mass $m_q^*$ and thus $\lambda_q$ are, in general, isospin
dependent \cite{prc69}. However we will consider an isospin independent
effective mass here for simplicity in this present study.
For the case of $m_q^* = m^*$ with $\lambda_q = \lambda$ 
(such as the case of $x_1 = x_2 = -1/2$), 
these equations become:
\begin{eqnarray}
 \mu_q(\rho,y,T) &=& T \ln\left[ \left(\frac{\lambda^3}{\gamma}\right)
      \left(\frac{\rho}{2} \pm (2y-1) \frac{\rho}{2} \right) \right]
    + \frac{T}{2\sqrt{2}} \left(\frac{\lambda^3}{\gamma}\right)
       \left(\frac{\rho}{2} \pm (2y-1)\frac{\rho}{2}\right)
           \nonumber \\  & &    \hspace{-2.0cm}
   + \frac{1}{4} \left[t_1 \left(1 + \frac{x_1}{2}\right)
       + t_2 \left(1 + \frac{x_2}{2}\right)\right] 
          \frac{3}{2} T \frac{2m^*}{\hbar^2} \left[\rho
        + \frac{\lambda^3}{2\sqrt{2} \gamma} 
           \left[1 + (2y-1)^2\right] \left(\frac{\rho}{2}\right)^2\right]
              \nonumber   \\  & &    \hspace{-2.0cm}
   - \frac{1}{4} \left[t_1 \left(\frac{1}{2} + x_1\right)
       - t_2 \left(\frac{1}{2} + x_2\right)\right] 
          \frac{3}{2} T \frac{2m^*}{\hbar^2}
          \left[\left(\frac{\rho}{2} \pm (2y-1)\frac{\rho}{2}\right) 
        + \frac{\lambda^3}{2^{5/2} \gamma} 
           \left(\frac{\rho}{2} \pm (2y-1)\frac{\rho}{2}\right)^2\right]
           \nonumber \\  & &
    + \frac{3}{4} t_0 \rho
    \mp \left(\frac{1}{2} + x_0\right) t_0 (2y-1)\left(\frac{\rho}{2}\right)
           \nonumber \\  & &
    + \frac{(\alpha+2)}{16} t_3 \rho^{\alpha+1}
    - \frac{1}{6} \left(\frac{1}{2} + x_3\right) t_3
          \left[\alpha (2y-1)^2 \left(\frac{\rho}{2}\right)^2
             \pm (2y-1)\left(\frac{\rho}{2}\right) \rho \right]
                 \rho^{\alpha-1}
              \nonumber   \\  & &
    + \frac{1}{4} C \left[\beta + 2 (1 \pm 1)\right] \rho^{\beta+1}
        + C \left[(\beta + 1 \pm 1) (2y-1) \left(\frac{\rho}{2}\right) \rho
                + \beta (2y-1)^2 \left(\frac{\rho}{2}\right)^2 \right]
            \rho^{\beta-1}
           \nonumber \\  & &
    + \eta C_s \rho^{\eta-1}   ,  \label{muryt}  \\
 P(\rho,y,T) &=& \left(\frac{5}{2} - \frac{3}{2} \frac{m^*}{m}\right) T \rho
    + \left(\frac{5}{2} - \frac{3}{2} \frac{m^*}{m}\right) \frac{T}{2\sqrt{2}} 
         \left(\frac{\lambda^3}{\gamma}\right) \left(\frac{\rho}{2}\right)^2
           \nonumber \\  & &
   + \frac{3}{8} t_0 \rho^2 + \frac{(\alpha+1)}{16} t_3 \rho^{\alpha+2}
   + \frac{(\beta+1)}{4} C \rho^{\beta+2} + (\eta - 1) C_s \rho^\eta
           \nonumber \\  & &   \hspace{-2.7cm}
     - \left[ t_0 \left(\frac{1}{2} + x_0\right)
        + \left(\frac{\alpha+1}{6}\right) t_3 \left(\frac{1}{2} + x_3\right)
             \rho^\alpha
        - \left(\frac{5}{2} - \frac{3}{2} \frac{m^*}{m}\right)
           \frac{T}{2\sqrt{2}} \left(\frac{\lambda^3}{\gamma}\right)
        - (\beta+1) C \rho^\beta
       \right] (2y-1)^2 \left(\frac{\rho}{2}\right)^2
           \nonumber \\ & & 
       + (\beta + 1) C \rho^{\beta+1} (2y-1) \left(\frac{\rho}{2}\right)
               \label{pryt}   ,   \\
 {\cal E}(\rho,y,T) &=& \frac{3}{2} T \rho
    + \frac{3}{8} t_0 \rho^2 + \frac{1}{16} t_3 \rho^{\alpha+2}
    + \frac{3}{2} \frac{T}{2\sqrt{2}} \left(\frac{\lambda^3}{\gamma}\right)
       \left(\frac{\rho}{2}\right)^2 + \frac{1}{4} C \rho^{\beta+2}
    + C_s \rho^\eta
           \nonumber \\  & &
     - \left[ t_0 \left(\frac{1}{2} + x_0\right)
        + \left(\frac{1}{6}\right) t_3 \left(\frac{1}{2} + x_3\right)
             \rho^\alpha
     - \frac{3}{2} \frac{kT}{2\sqrt{2}} \left(\frac{\lambda^3}{\gamma}\right)
        - C \rho^\beta
       \right] (2y-1)^2 \left(\frac{\rho}{2}\right)^2
           \nonumber \\ & &
       + C \rho^{\beta+1} (2y-1) \left(\frac{\rho}{2}\right)
               \label{eryt}    ,   \\
 T\calS(\rho,y,T) &=& T \rho \left[\frac{5}{2}
        - y \ln\left(\frac{\lambda^3}{\gamma} y\rho\right)
        - (1-y) \ln\left(\frac{\lambda^3}{\gamma} (1-y)\rho\right) \right]
           \nonumber \\ & &
      + \frac{T}{2\sqrt{2}} \left(\frac{\lambda^3}{\gamma}\right)
        \frac{[1 + (2y-1)^2]}{2} \left(\frac{\rho}{2}\right)^2
               \label{tsryt}
\end{eqnarray}
Here, for the proton density ($\rho_p$) and neutron density ($\rho_n$),
we defined, the isoscalar density $\rho$, isovector density $\rho_3$,
proton fraction $y$ and related quantities by 
\begin{eqnarray}
& \rho = \rho_p + \rho_n  ,  \hspace{5ex}
 \rho_3 = \rho_p - \rho_n = (2 y - 1) \rho  ,
    \hspace{5ex} y = \rho_p/\rho  ,
    \nonumber \\
& \rho_p = \frac{1}{2} (\rho + \rho_3) = y \rho  ,   \hspace{5ex}
 \rho_n = \frac{1}{2} (\rho - \rho_3) = (1-y) \rho  ,   \\
& \sum_q \rho_q^2 = \frac{1}{2} (\rho^2 + \rho_3^2)
     = \frac{[1 + (2y-1)^2]}{2} \rho^2 = [1 + 2y(y-1)]\rho^2  ,  \nonumber \\
& \sum_q \rho_q^3 = \frac{1}{4} \rho(\rho^2 + 3\rho_3^2)
     = \frac{[1 + 3(2y-1)^2]}{4} \rho^3 = [1 + 3y(y-1)]\rho^3  \nonumber
\end{eqnarray}
The $\pm$ in $\mu_q$ stands $+$ for $q=$proton and $-$ for neutron.

At fixed $T$ and $P$, only one of either $\rho$ or $y$ is
the independent variable.
Thus observables such as $P$, ${\cal E}/\rho$, $\calS/\rho$
may have a discontinuity in $T$ or $y$
when $\left(\frac{\partial\rho}{\partial T}\right)_{y,P}$
or $\left(\frac{\partial\rho}{\partial y}\right)_{T,P}$
diverges.
We can study the behavior of thermodynamic quantities at a fixed $P$
using $d P = 0$ from Eq.(\ref{pryt}),
\begin{eqnarray}
 dP 
  &=& \left\{ \left(\frac{5}{2} - \frac{3}{2} \frac{m^*}{m}\right) \left[ \rho
    - \frac{1}{2} \frac{1}{2\sqrt{2}} \left(\frac{\lambda^3}{\gamma}\right)
       \left(\frac{\rho}{2}\right)^2
    - \frac{1}{2} \frac{1}{2\sqrt{2}} \left(\frac{\lambda^3}{\gamma}\right)
       \left(2y-1\right)^2 \left(\frac{\rho}{2}\right)^2
             \right] \right\} dT
           \nonumber \\ & &
    + \left\{
       \left[\frac{5}{2} - \frac{3}{2} \left(\frac{m^*}{m}\right)^2\right] T 
            + \frac{3}{4} t_0 \rho 
            + \frac{(\alpha+2)(\alpha+1)}{16} t_3 \rho^{\alpha+1}
        \right.   \nonumber \\ & &
    + \left[\frac{35}{8} - \frac{15}{4} \frac{m^*}{m}
           + \frac{3}{8} \left(\frac{m^*}{m}\right)^2 \right] 
       \frac{T}{2\sqrt{2}} \left(\frac{\lambda^3}{\gamma}\right)
       \left(\frac{\rho}{2}\right) 
      + \frac{(\beta+2)(\beta+1)}{4} C \rho^{\beta+1}
      + \eta(\eta-1) C_s \rho^{\eta-1}     
           \nonumber \\  & &
     - \left[ t_0 \left(\frac{1}{2} + x_0\right)
        + \left(\frac{\alpha+2}{2}\right) \left(\frac{\alpha+1}{6}\right)
              t_3 \left(\frac{1}{2} + x_3\right) \rho^\alpha
        - \left(\frac{\beta+2}{2}\right) (\beta+1) C \rho^\beta
        \right.   \nonumber \\ & &   \left.
        - \left(\frac{35}{8} - \frac{15}{4} \frac{m^*}{m}
               + \frac{3}{8} \left(\frac{m^*}{m}\right)^2 \right)
           \frac{T}{2\sqrt{2}} \left(\frac{\lambda^3}{\gamma}\right)
        \right] (2y-1)^2 \left(\frac{\rho}{2}\right) 
              \nonumber \\ & &      \left.
     + (\beta+2) (\beta + 1) C \rho^{\beta} (2y-1)
            \left(\frac{\rho}{2}\right)  \right\} d\rho
           \nonumber \\
   &-& \left\{ \left[ t_0 \left(\frac{1}{2} + x_0\right)
        + \left(\frac{\alpha+1}{6}\right) t_3 \left(\frac{1}{2} + x_3\right)
             \rho^\alpha
        - (\beta+1) C \rho^\beta
        \right. \right.   \nonumber \\ & &  \left.   \left.
        - \left(\frac{5}{2} - \frac{3}{2} \frac{m^*}{m}\right) 
            \frac{T}{2\sqrt{2}} \left(\frac{\lambda^3}{\gamma}\right)
       \right] (2y-1) \left(\frac{\rho}{2}\right)^2    
       - (\beta + 1) C \rho^{\beta} \left(\frac{\rho}{2}\right)^2
             \right\} 4 dy
\end{eqnarray}
This equation gives $y_E(\rho)$ where both $\partial P/\partial y = 0$ 
and $\partial\rho/\partial y = 0$, 
\begin{eqnarray}
 y_E(\rho) &=& \frac{1}{2} 
    + \frac{1}{2} \frac{(\beta+1) C \rho^\beta}  {
    \left[ t_0 \left(\frac{1}{2} + x_0\right)
        + \left(\frac{\alpha+1}{6}\right) t_3 \left(\frac{1}{2} + x_3\right)
             \rho^\alpha
        - (\beta+1) C \rho^\beta
        - \left(\frac{5}{2} - \frac{3}{2} \frac{m^*}{m}\right)
            \frac{T}{2\sqrt{2}} \left(\frac{\lambda^3}{\gamma}\right)
       \right] } .    \label{yerho}
\end{eqnarray}
The $y_E(\rho)$ is indepenent of $\rho$ for momentum independent Skyrme
interaction with $x_3 = -1/2$ and $\beta = 0$
as considered in Ref.\cite{prc63,plb580}.
The $x_3$ term and the density dependent effective mass for a momentum 
dependent Skyrm force introduce a small $\rho$-dependence in $y_E$.
Eq.(\ref{pryt}) shows that, for $\rho$-dependent $y_E$,
the $P(\rho)$ curve for different values of $y$ at fixed $T$ may cross
at some $\rho$. Moreover, the minimum pressure for a given $T$
and $\rho$ (i.e., $(\partial P/\partial y)_{\rho,T} = 0$) occurs
at $y = y_E(\rho) \ne 0.5$ due to Coulomb effect.
These results were not seen in Ref.\cite{serot}.
At $y_E$, the pressure of the coexistence curve is minimum and 
the liquid and gas phases have the same proton fraction $y_E$.
The condition $\partial P/\partial y = 0$ determines the equal
fraction point $y_E$.

\section{Applications to Nuclear Mechanical and Chemical Instability
and the Liquid-Gas Phase Transition}  \label{rsltsec}

\subsection{Mechanical and Chemical Instability}

%
%
\begin{figure}
\includegraphics[width=5.0in]{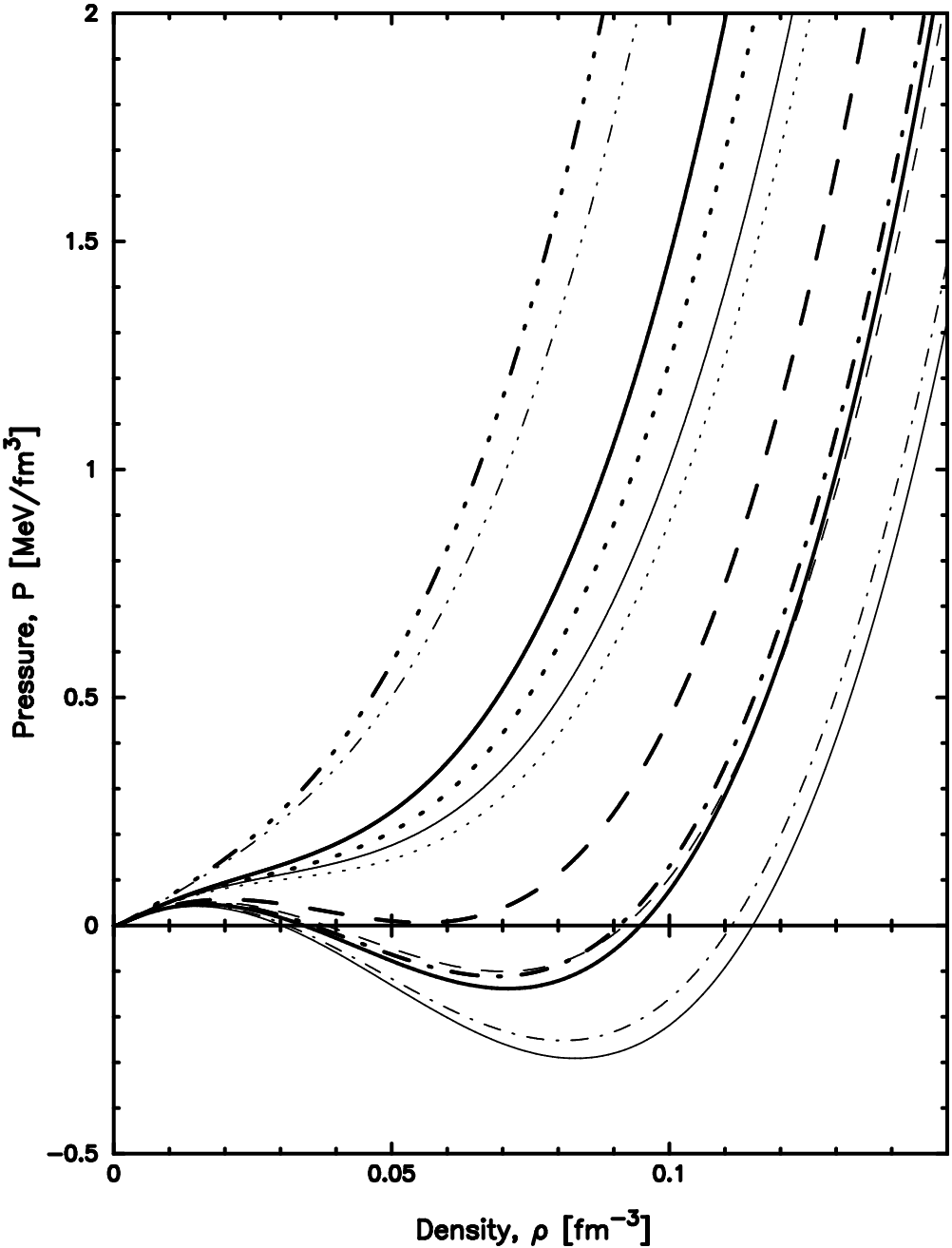}
\caption{Pressure $P(\rho)$ versus $\rho$ at $T = 10$ MeV for various proton
fraction $y$.
The upper solid curve is for $y = 0$, dashed curve for $y = 0.2$,
dash-dotted curve for $y = 0.5$, dotted curve for $y = 0.8$,
and dash-dot-dot-dotted curve for $y = 1$. 
The thick curves are for the momentum dependent Skyrme force
and the thin curves are for the momentum independent Skyrme force.
The lower thick solid curve is for $y = y_E(\rho)$ of Eq.(\ref{yerho})
($y_E = 0.4106 \sim 0.4214$ for $\rho = 0 \sim 0.15$ fm$^{-3}$)
with momentum dependent Skyrme force and the lower thin solid curve is 
for $y = y_E = 0.41057$ with momentum independent Skyrme force.
 }   \label{fig1}
\end{figure}

We now use the results to discuss features of the instability of nuclei,
both mechanical and chemical, and the liquid-gas phase transition.
The region of mechanical instability is determined by the condition
\begin{eqnarray}
 \left.\frac{dP}{d\rho}\right|_{y,T} = 0 
\end{eqnarray}
Fig.\ref{fig1} shows the behavior of the pressure $P(\rho, y, T)$ as 
a function of $\rho$ for several values of the proton fraction $y$.
All curves are at $T = 10$ MeV.
The range of $y$ is from $y = 0$, or pure neutron matter,
to $y = 1$, or pure proton systems.
The point $y = 1/2$ corresponds to symmetric systems.
Without a Coulomb interaction results would be symmetric about
the point $y = 1/2$ which would also be the point of equal concentration in
a liquid/gas phase coexistence. Including a Coulomb interaction shift
the equal concentraion point to a proton fraction 
of $y = y_E(\rho) \sim 0.415$ with a momentum 
dependence included in the interaction and 
to $y = y_E = 0.41057$ without a momentum dependence.
The $y = y_E$ curve for both the momentum dependent and independent cases
has the lowest pressure versus density dependence,
i.e., the lowest $P$ for a given value of $\rho$ at a given $T$.
A higher or lower $y$ raises the pressure at a given density.
Both a momentum dependent Skyrme interaction and a momentum independent
Skyrme interaction results are shown 
for several values of $y$ and they are distinguished by the thickness of
the lines as described in the figure caption.
The momentum dependence increases the pressure in the range shown
(thick lines compared to thin lines) and introduces 
the density dependence of $y_E(\rho)$ given by Eq.(\ref{yerho}).
The mechanical instability densities for each $y$ curve at $T = 10$ MeV
are the points where the $P(\rho, y, T)$ curve has zero slope,
$dP/d\rho|_{y,T} = 0$.
The total region of mechanical instability is obtained by a similar
calculation of $P(\rho, y, T)$ at different $T$.
For a one component system or a symmetric system the mechanical instability 
region is a curve somewhat similar to an inverted parabola with its peak
at the critical point.
Allowing for systems with different values of $y$ gives a two dimensional
boundary surface for the mechanical instability region.
The intersection of the surface with different $y$ planes gives the one 
dimensional boundary curve or line of mechanical instability for each
corresponding value of $y$.

%
%
\begin{figure}
\includegraphics[width=5.0in]{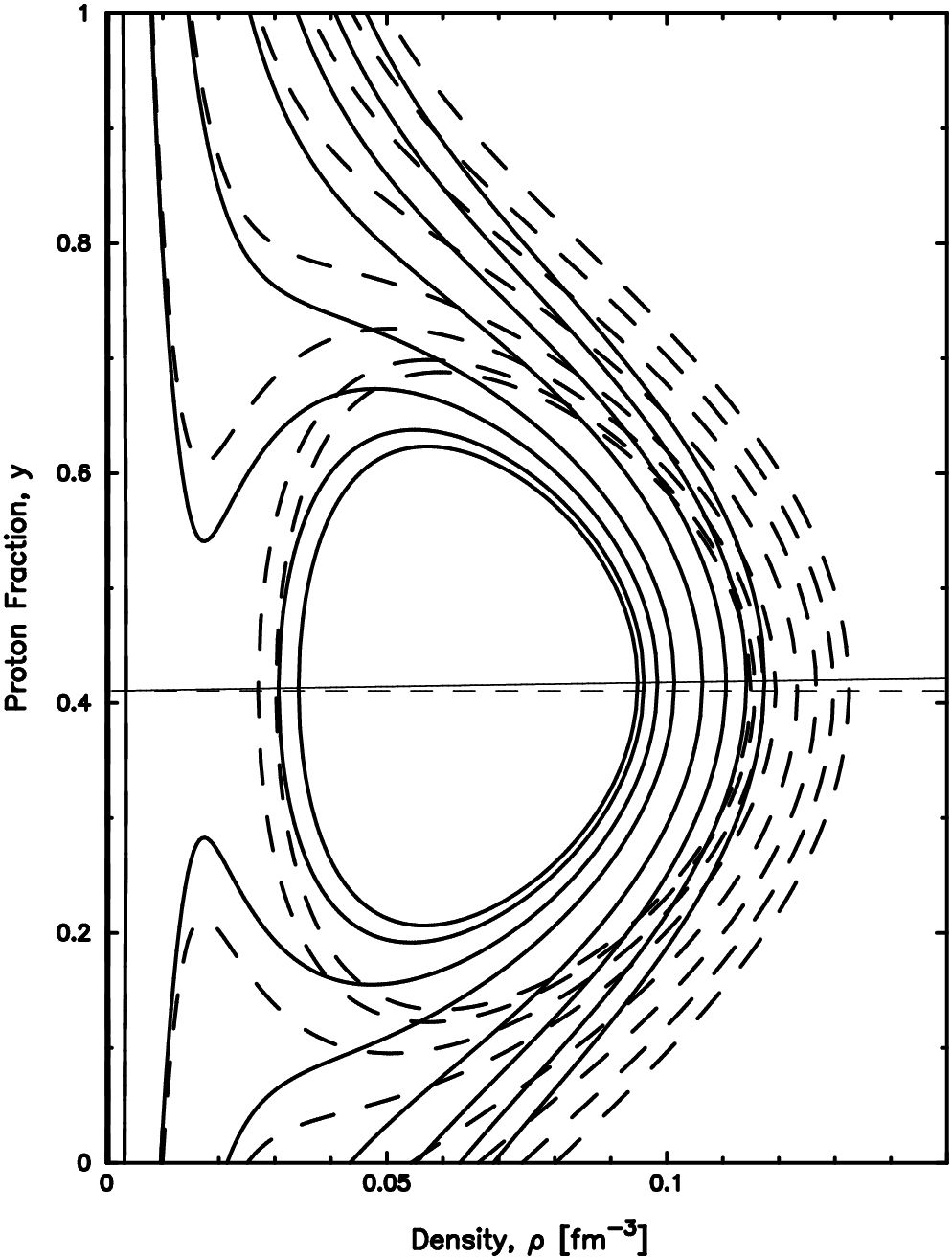}
\caption{Proton fraction $y(\rho)$ for $P = 0$, 0.015, 0.05, 0.1, 0.2, 0.3,
0.4, and 0.5 MeV/fm$^3$ from inside to outside at $T = 10$ MeV.
The solid curves are for the momentum dependent Skyme force
and the dashed curves are for the momentum independent Skyrme force.
The thin straight lines are $y_E(\rho)$ for the corresponding force.
 }    \label{fig2}
\end{figure}

Fig.\ref{fig2} shows the proton fraction $y$ versus the density $\rho$ for 
different fixed values of the pressure at a fixed temperature of $T = 10$ MeV.
The loops and curves are determined by solving $P(\rho, y, T) = P$ for
the values of $P$ listed in the figure caption and at the
temperature $T = 10$ MeV.
Fig.2 is obtained from Fig.1 by drawing a horizontal line and 
looking at the points where the horizontal line intersects the 
set of $P(\rho, y, T)$ curves. 
This intersection can be at one, two or three points. 
Besides the inner most closed loop ($P = 0$) shown in Fig.\ref{fig2}, 
a vertical line exists at $\rho = 0$ for $P = 0$ for all $y = 0 \sim 1$.
Similarly, for the second inner closed loop at $P = 0.015$ MeV/fm$^3$, 
a nearly parallel vertical line is present at very low density. 
The right most point on each curve and the left most point on a closed loop
with $d\rho/dy|_{P,T} = 0$ are at the point of equal concentration $y_E$. 
Also shown are two thin lines for $y_E(\rho)$. 
The dashed thin line is at $y_E = 0.41057$ and is horizontal or density 
independent and corresponds to the momentum independent interaction. 
The solid thin line is nearly horizontal with a slight density dependence 
and has $y_E(\rho) = 0.4106 \sim 0.4214$ for $\rho = 0 \sim 0.15$ fm$^{-3}$. 
Horizontal turning points on each curve occur at $dy/d\rho|_{P,T} = 0$. 
For each $T$, there is a curve $P(\rho)|_{y,T}$ with an inflection point
for a particular $y$ which we call $y_I$.
At the pressure $P = P(\rho, y_I, T)$, the closed loop in Fig.\ref{fig2}
just breaks at the point of $y = y_E$ on the left low density side and 
creates two new horizontal turning points with $\partial y/\partial\rho = 0$.
Fig.2 also shows the result that a momentum independent force has 
closed loops outside those of a momentum dependent force  
and open curves to the right of those of a momentum dependent force
with the same pressure $P$.

\begin{figure}
\includegraphics[width=5.0in]{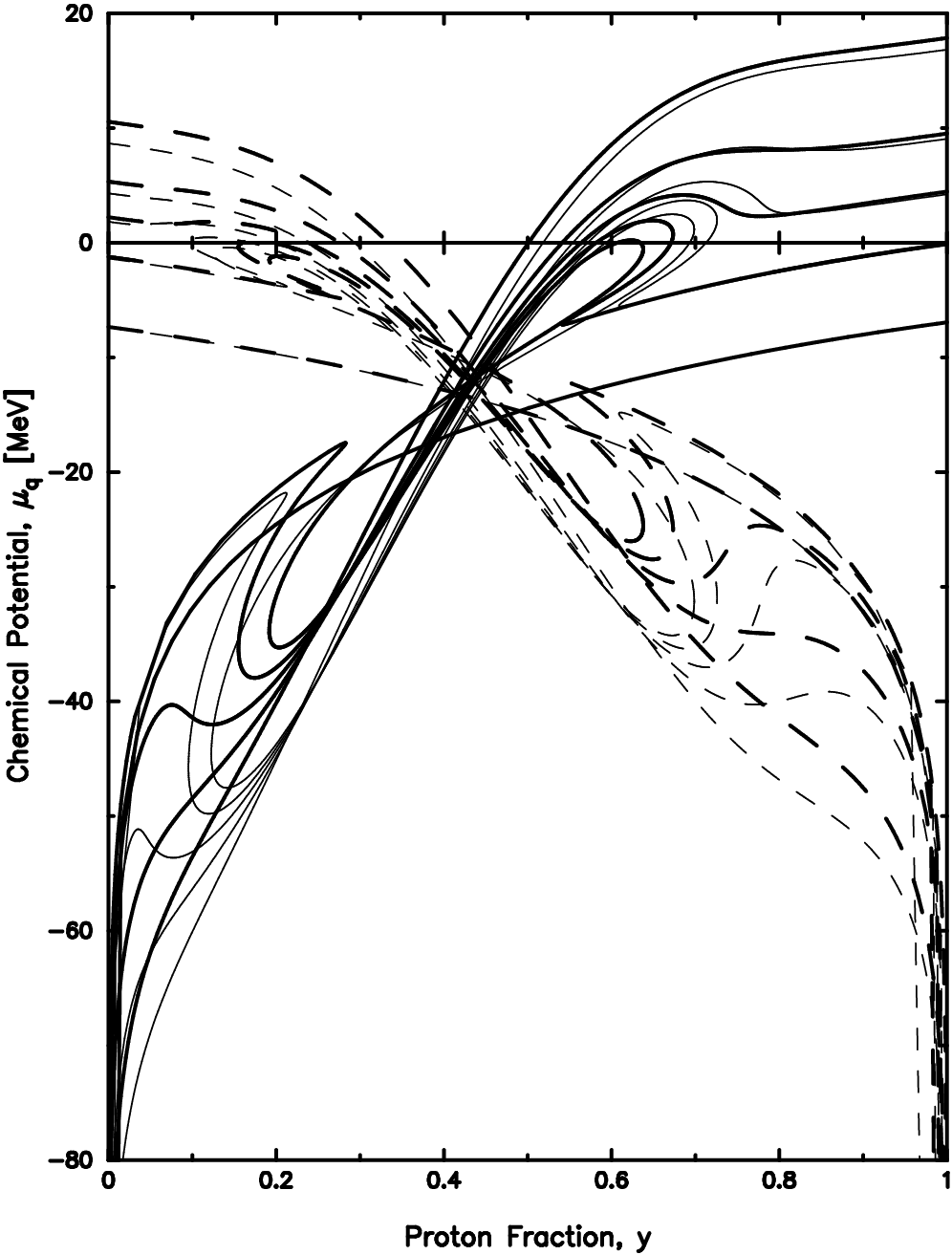}
\caption{Chemical potential $\mu_p(y)$ and $\mu_n(y)$
for $P = 0.015$, 0.05, 0.1, 0.2, and 0.5 from top to bottom curve for protons
(solid curve) and from bottom to top curve for neutrons (dashed curve)
at $T = 10$ MeV.
Thick curves are for momentum dependent Skyrme force and
the thin curves for momentum independent Skyrme force.
 }  \label{fig3}
\end{figure}

\begin{figure}
\includegraphics[width=5.0in]{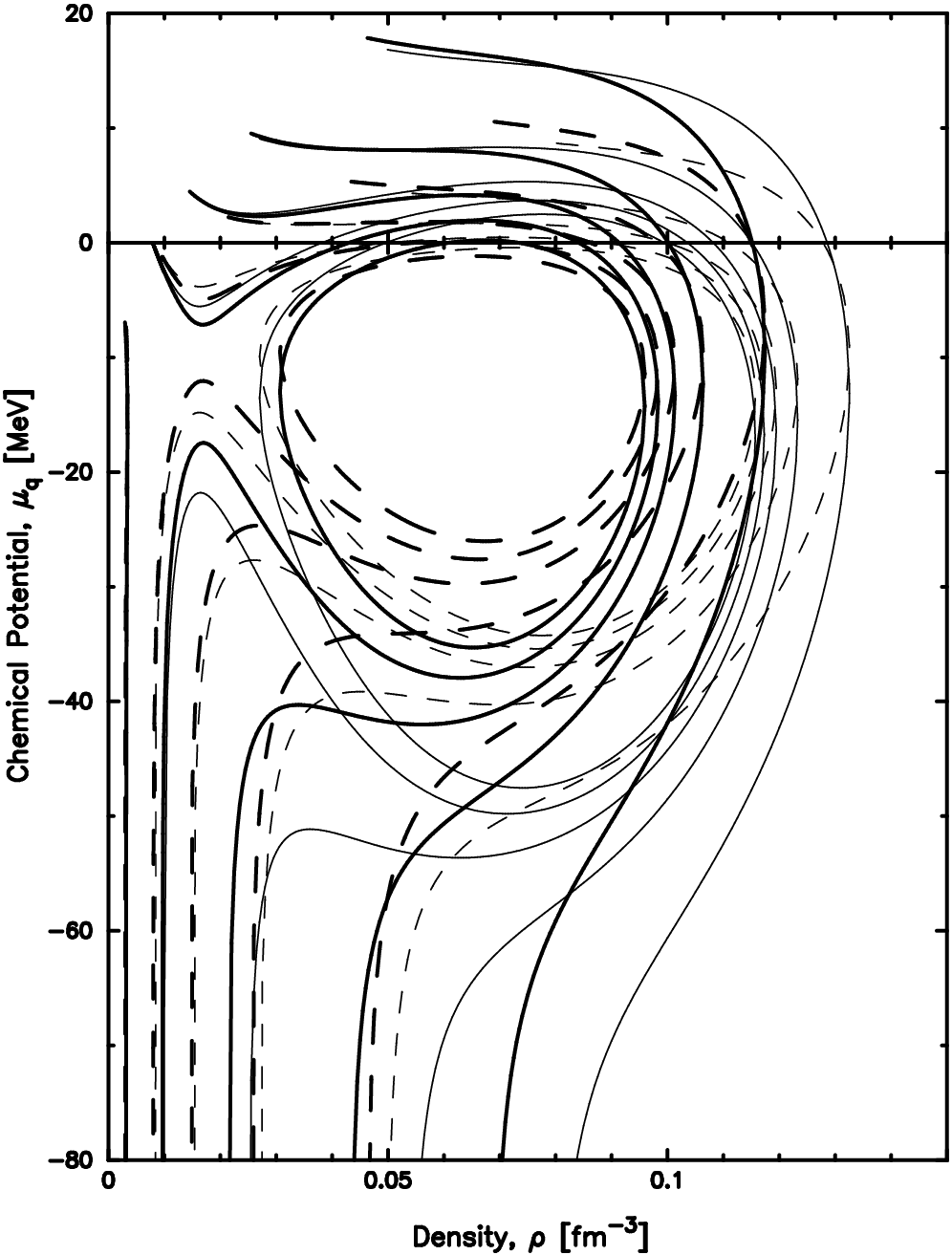}
\caption{Same as Fig.\ref{fig3} but for $\mu_p(\rho)$ and $\mu_n(\rho)$ 
versus $\rho$.
 }    \label{fig4}
\end{figure}

The region of chemical instability (spinodal in $\mu(y)|_{P,T}$) is
determined by the condition
\begin{eqnarray}
 \left. \frac{d\mu_q}{dy}\right|_{P,T} = 0
\end{eqnarray}
for each component $q = p$ or $n$.
These conditions for either protons or neutrons give the same relation since
\begin{eqnarray}
 y d\mu_p + (1-y) d\mu_n = \frac{1}{\rho} dP   \label{dmupn}
\end{eqnarray}
This general condition will be used later in our discussion of results
given in various figures. The result is also useful for checking
numerical results.
The chemical instability condition can be rewritten in terms of derivatives
of the chemical potential and pressure with respect to the density
variable $\rho$ and proton fraction $y$.
Namely, the chemical instability condition can be obtained from the
following relation \cite{plb580}
\begin{eqnarray}
 \left. \frac{dP}{d\rho}\right|_{y,T} 
       \left. \frac{d\mu_q}{dy}\right|_{\rho,T} 
  &=& \left.\frac{dP}{dy}\right|_{\rho,T} \left.\frac{d\mu_q}{d\rho}\right|_{y,T}
\end{eqnarray}
The expressions developed for the proton and neutron chemical potentials
are functions of the variables ($\rho$, $y$, $T$). 
The equation of state $P(\rho, y, T)$ can then be used to find their
behaviors in terms of ($y$, $P$, $T$) or ($\rho$, $P$, $T$). The behaviors 
with $y$ of the proton chemical potential $\mu_p(\rho, P, T) \to \mu_p(y)$ and
neutron chemical potential $\mu_n(\rho, P, T) \to \mu_n(y)$ at various
values of the pressure $P$ and at a fixed temperature $T = 10$ MeV are
shown in Fig.\ref{fig3}. 
The chemical instability region boundaries are determined by the points 
where the slope of each chemical potential with respect to $y$ is zero. 
Further discussion of the chemical spinodal line is given 
in the next subsection. 
The behaviors of the proton 
chemical potential $\mu_p(\rho, P, T) \to \mu_p(\rho)$ and
neutron chemical potential $\mu_n(\rho, P, T) \to \mu_n(\rho)$ 
with density $\rho$ at various fixed values of the pressure $P$ and at
a fixed temperature $T = 10$ MeV are shown in Fig.\ref{fig4}.
Fig.\ref{fig2} and Fig.\ref{fig4} show some similarities in the behavior 
of the plotted quantities, i.e., inner closed loops at low pressure,
to outer curves that almost form closed loops with increasing pressure,
to open curves with further increases in pressures.

%
%
%

%
%

\subsection{Liquid-Gas Phase Transition and the Coexistene Curve}

\begin{figure}
\includegraphics[width=5.0in]{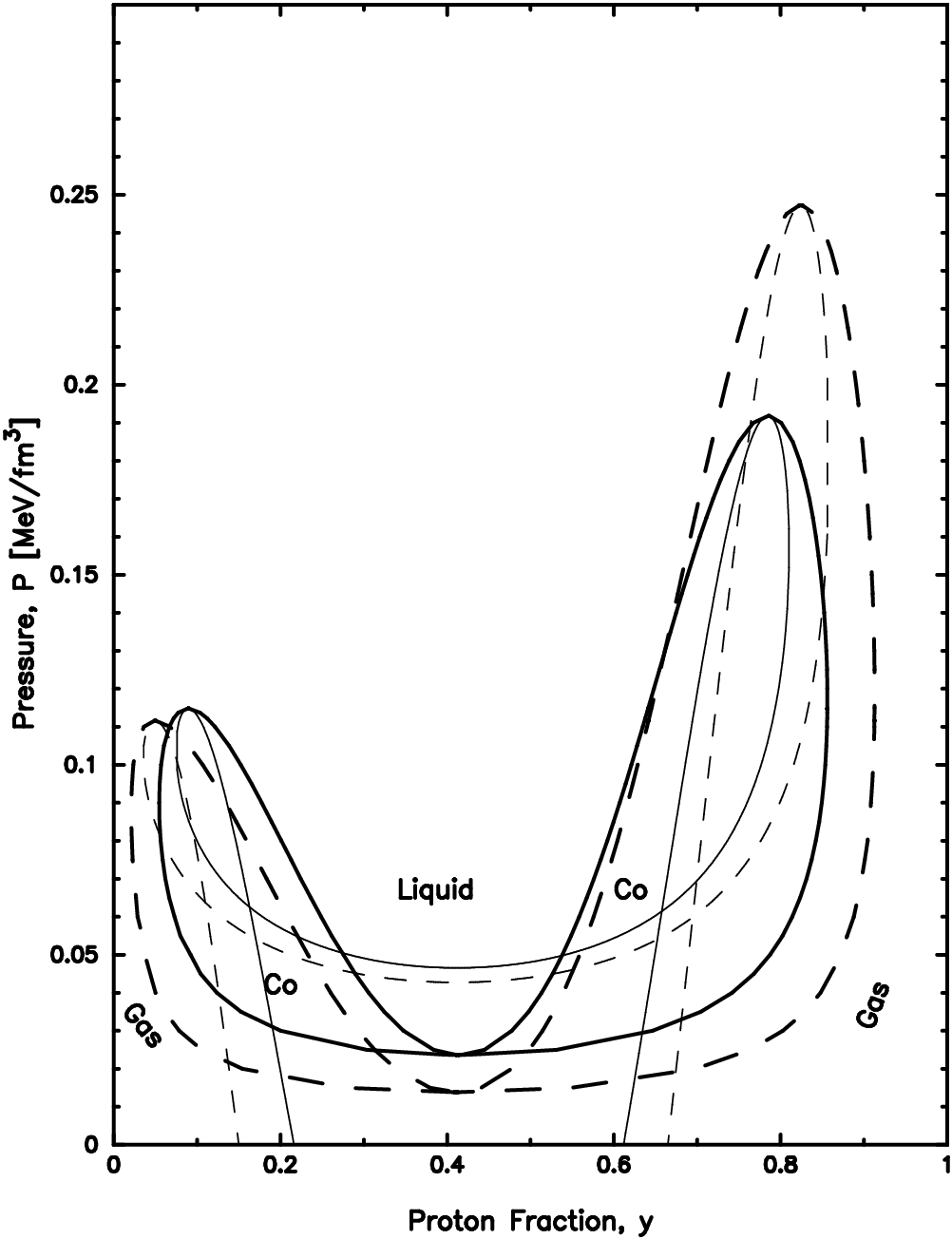}
\caption{Pressure $P$ versus proton fraction $y$ for coexistence loop
(thick curves) at $T = 10$ MeV. 
The solid curve is for the momentum dependent Skyrme force
and the dashed curve is for momentum independent Skyrme force.
The thin curves are the chemical instability boundary curves for 
each case of Skyrme interaction respectively.
For both momentum dependent and momentum independent cases
the maximum of the chemical instability loop and the coexistence
loop occur at the same point where the curves are tangent to
each other as discussed in the text.
The point of equal concentration is $y_E \sim 0.415$ for momentum
dependent case and $y_E = 0.41057$ for momentum independent case.
 }  \label{fig5}
\end{figure}

\begin{figure}
\includegraphics[width=5.0in]{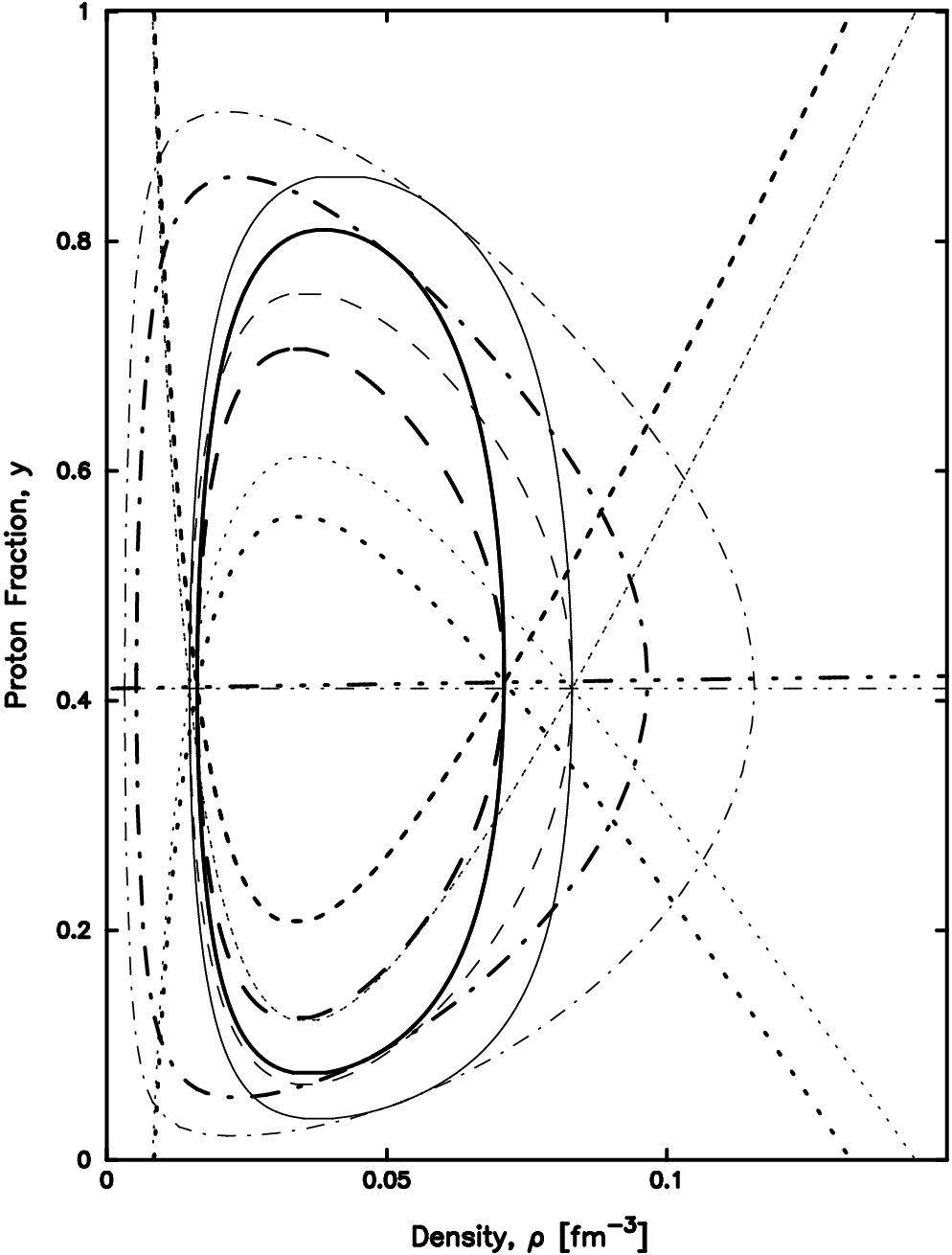}
\caption{Figure shows the coexistence curves (dash-dotted line),
chemical instability boundary curves (solid line) and 
mechanical instability boundary curves (dashed line) at $T = 10$ MeV. 
Also shown are the $\partial\mu_q/\partial\rho = 0$ 
curves for proton (dotted line) and for neutron (short dash line) 
at $T = 10$ MeV.
The dash-dot-dot-dotted line is for $y_E(\rho)$.
The thick lines are for momentum dependent Skyrme force and
the thin lines are for momentum independent Skyrme force.
The momentum dependent loops are inside the momentum independent loops.
  }  \label{fig6}
\end{figure}

\begin{figure}
\includegraphics[width=5.0in]{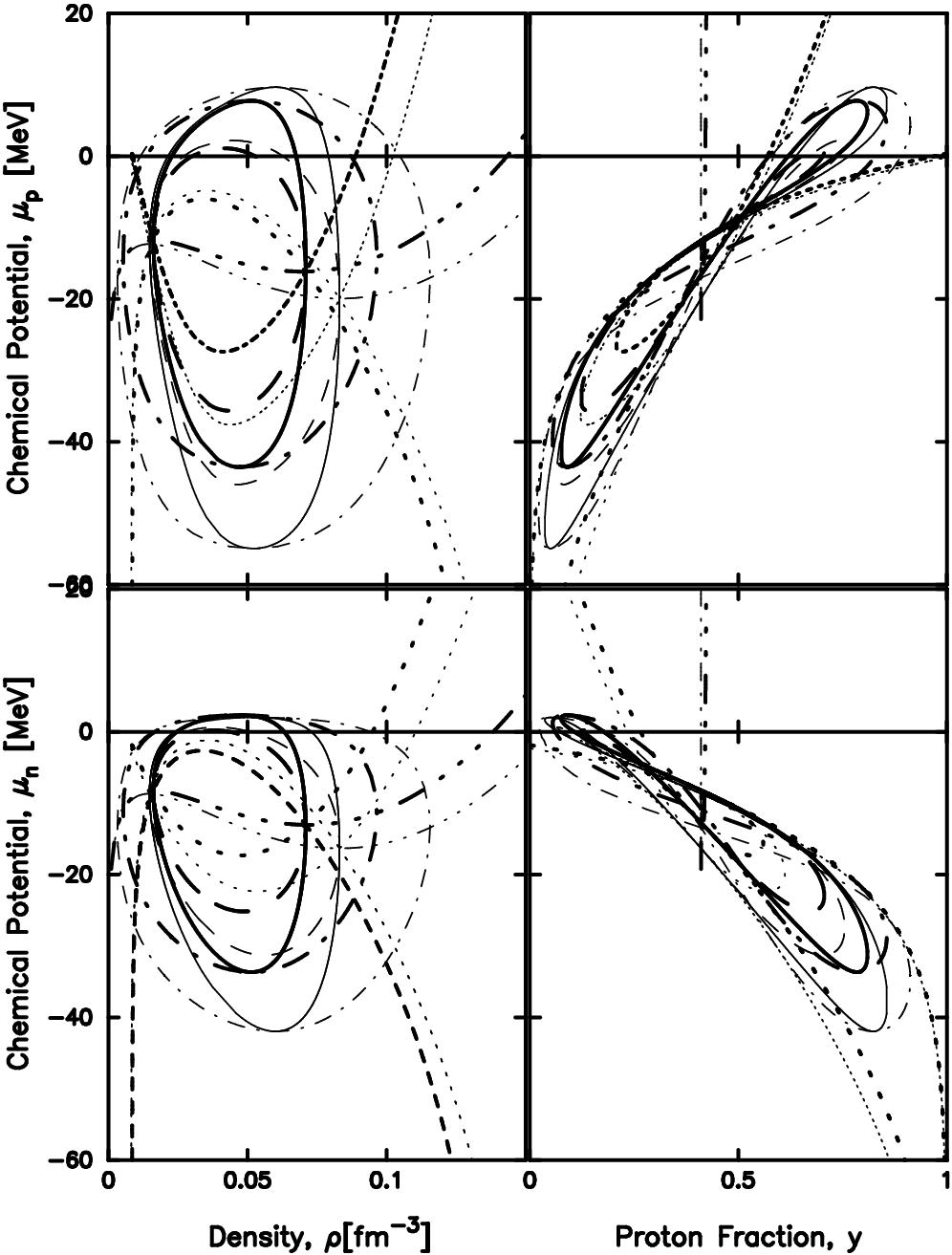}
\caption{Chemical potential $\mu_p$ (upper panel) 
and $\mu_n$ (lower panel) for various
boundary curves at $T=10$ MeV. 
The curves are same as in Fig.\ref{fig6}.
  }  \label{fig7}
\end{figure}

\begin{figure}
\includegraphics[width=5.0in]{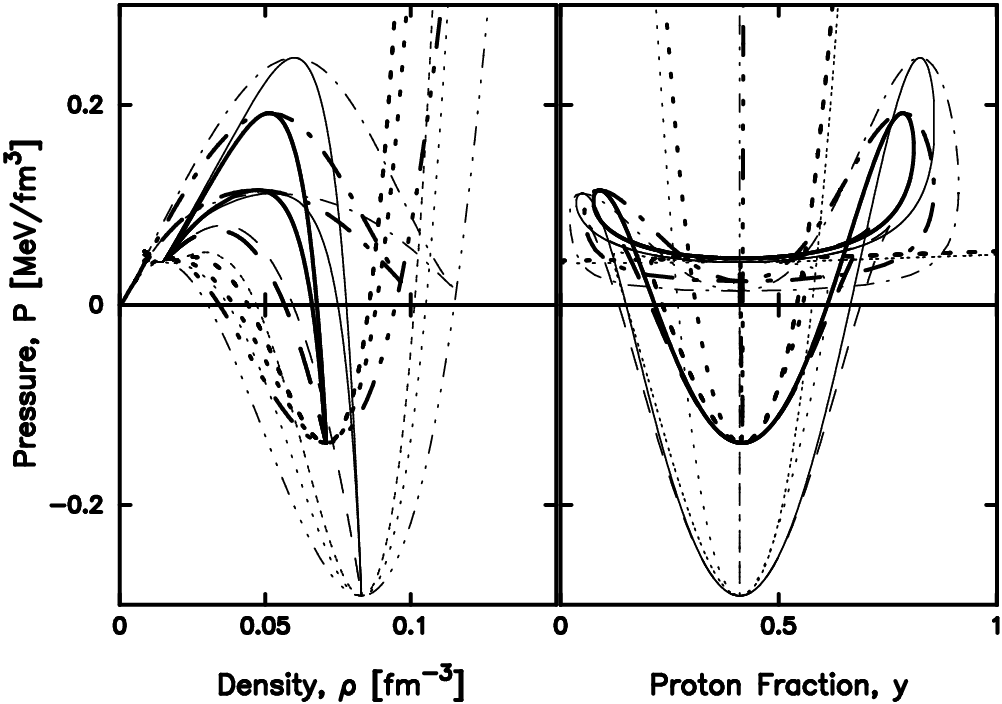}
\caption{Pressure $P$ for various curves.
The curves are same as in Fig.\ref{fig6}.
  }  \label{fig8}
\end{figure}

\begin{figure}
\includegraphics[width=5.0in]{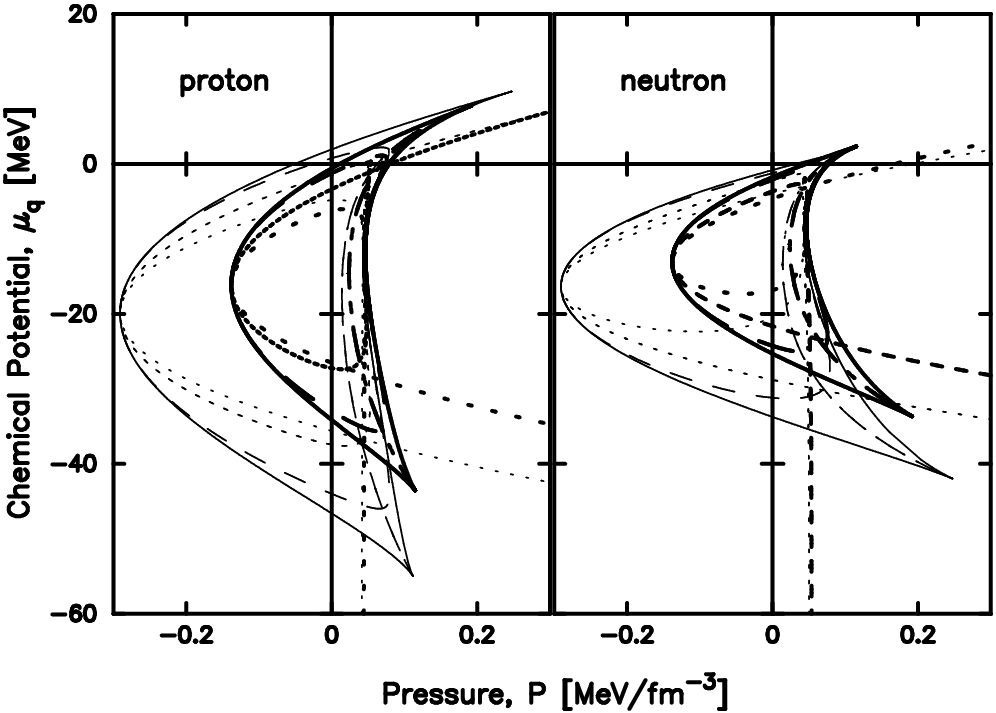}
\caption{Chemical pressure $\mu_q$ versus pressure $P$
for various boundary curves. 
The left pannel is for proton $\mu_p$
and the right pannel is for neutron $\mu_n$.
The curves are same as in Fig.\ref{fig6}.
  }  \label{fig9}
\end{figure}

For a one component system the coexistence curve is a line obtained by
the familiar Maxwell construction as already noted. For a two component 
system the coexistence region is a surface obtained as follows.
The condition for coexistence between the two phases requires the proton
chemical potentials to be the same in two phases and, similarly,
the neutron chemical potentials must be the same in the two phases at
a given pressure and temperature.
Note that the proton fraction need not be the same in each of the two phases.
In fact, the liquid phase should be a more symmetric system than the gas 
phase because of the symmetry potential as seen in Refs.\cite{prc63,plb580}.
Figs.\ref{fig5}-\ref{fig9} show features of the coexistence curves
together with the mechanical and chemical instability curves.

The condition of phase coexistence corresponds to a rectangular box
geometrical construction in the chemical potential plots of Fig.\ref{fig3}
or of Fig.\ref{fig4}.
Namely, the chemical potential equality condition 
 $\mu_p(y_1, P, T) = \mu_p(y_2, P, T)$ 
and $\mu_n(y_1, P, T) = \mu_n(y_2, P, T)$ leads
to a rectangular box in Fig.\ref{fig3} with vertical sides connecting 
the $\mu_p(y_1, P, T)$ to the $\mu_n(y_1, P, T)$ for side 1 
and the $\mu_p(y_2, P, T)$ to the $\mu_n(y_2, P, T)$ for side 2.
The horizontal sides are the chemical potential equality
conditions at $y_1$ and $y_2$ for neutrons and for protons.
The rectangular box shrinks in its horizontal direction in $\mu_q$-$y$ plots
as the point of equal concentration, where the liquid and gas phases have
the same proton fraction, is approached (the lowest point
of the coexistence curve in Fig.\ref{fig5}).

%
%

Fig.\ref{fig5} shows various features of the coexistence region
in pressure versus proton fraction.
The coexistence region are the dark thicker solid line
for a momentum dependent force and the dark thicker dashed
line for a momentum independent force. Also shown are associated
chemical instability regions as a thinner solid line, and thinner dashed line.
The calculations are done at a temperature of 10 MeV.
For a two component system, the coexistence and instability regions are
two dimensional surfaces in pressure, temperature and proton fraction
as mentioned above.
The pressure-proton fraction behavior shown is a consequence of
cutting these surfaces with a constant temperature plane.
The result at $T = 10$ MeV are the loops shown.
Other temperatures can be obtained in a similar fashion.
For a momentum independent force the chemical instability
region basically lies inside the coexistence curve and peaks
at the top of the coexistence loop, the critical points.
The condition $dP/dy|_T = 0$ with $d^2P/dy^2|_T < 0$ 
gives a critical point on the coexistence curve 
and the condition $dy/dP|_T = 0$ gives the point with maximal asymmetry
at the left and right most points of coexistence curve.
The proton rich $y \ge y_E$ and neutron rich $y \le y_E$ loops 
are very asymmetric because of the Coulomb interaction. 
The inclusion of velocity or momentum dependent interactions 
leads to further modification of the coexistence curve and
chemical instability curves.
This modification is easily seen in the figure by comparing the dashed 
momentum independent curves with the solid momentum dependent case.
The figure shows that the momentum dependent interaction that was used
has a larger effect on the asymmetric proton rich loop ($y > y_E$)
significantly reducing its maximum pressure.
The maximum of the neutron rich loop ($y < y_E$) remains somewhat
unchanged with a small increase.
Another effect is to shift the two loops inward toward the
equal concentration point $y_E$.
A third effect is to shift the lowest pressure point, which occures
at the equal concentration $y_E$, upward 
with the value of $y_E$ nearly unchanged.
Finally, it should be noted that the peaks of the coexistence and
chemical instability curves are at the same point 
where the curves are tangent to each other.
We see no indication of a truncation effect in our model 
where the coexistence curve intersects the chemical instability
curve before reaching the peak critical point.
A truncation effect gives a limiting pressure (below the maximum
pressure of the chemical instability curve) above which
a liquid-gas phase transition cannot take place \cite{plb650}.

%
%

Fig.\ref{fig6} shows plots in $y$ versus $\rho$ of phase coexistence curves, 
instability boundary loops for both chemical and mechanical instability,
and features of $\partial\mu_q/\partial\rho|_{y,T} = 0$ for proton and neutrons.
The thin curves are for a momentum independent interaction and
the thick curves are for a momentum dependent interaction.
The calculations are done at a fixed temperature of 10 MeV.
Some features common to both cases are as follows.
The mechanical and chemical instability boundary curves are
closed loops with the mechanical loop (dashed line) inside
the chemical instability loop (solid line).
These two loops touch at $y_E$, the dash-dot-dot-dotted line.
The $y_E(\rho)$ increases slightly with $\rho$ for a momentum 
dependent interaction while it is constant (horizontal) for a 
momentum independent interaction. 
The $\rho$ dependence of $y_E(\rho)$ come from the $\rho$ dependence
of the effective mass and also from the $x_3$ term
as can be seen in Eq.(\ref{yerho}). 
Also intersecting at these same points 
are $\partial\mu_p/\partial\rho = 0$ and $\partial\mu_n/\partial\rho = 0$. 
Different features and behaviors exist between the two cases.
The momentum dependent case (thick curves) has behaviors that are 
compressed in these $y$-$\rho$ plots. 
The coexistence curves have a different quantitative but similar qualitative
behavior between the two cases. 
The coexistence loop (dash-dotted line) is outside the other 
two loops and tangent to chemical instability loop at two points.
These two points are the critical points of low and high $y$
which are shown in Fig.\ref{fig5} where the two loops touch at 
the peak of each loop.
Comparing the two cases quantitatively, we see a compression of the results
of the momentum dependent case (thick curves) with respect to the
results of the momentum independent case (thin curves).
The thick loops are inside of thin loops.

%
%
%

%
%

Fig.\ref{fig7} and Fig.\ref{fig8} show chemical potentials for both
proton $\mu_p$ and neutron $\mu_n$ and pressure $P$ along the various
curves of coexistence and chemical and mechanical instabilities.
Curves in Fig.\ref{fig7} illustrate the
behavior of each chemical potential with density on the
left panel and proton fraction on the right panel.
Curves in Fig.\ref{fig8} are pressure versus density on the left
side and pressure versus proton fraction on the right.
The separate pressure-proton fraction behaviors in 
Fig.\ref{fig8} were already shown in Fig.\ref{fig5}, but now these two
figures contain additional plotted quantities which are 
the $\partial\mu_q/\partial\rho|_{y,T} = 0$ curves.
The chemical potential density curves in Fig.\ref{fig7}  
have features similar to those discussed in Fig.\ref{fig6}.
Both momentum dependent and independent cases of Fig.\ref{fig7} 
and Fig.\ref{fig8} shows tangent points of
the solid line and dash-dotted line. 
Also seen in these figures are the compression or shrinking of various curves 
for momentum dependent case with respect to the momentum independent case.
The thick loops are inside of thin loops.
Fig.\ref{fig5} shows that the momentum dependent interaction leaves the
point of equal concentraion nearly unchanged
i.e., from $y = 0.4106$ to $y \approx 0.415$.

%
%

Fig.\ref{fig9} shows the behavior of the boundary curves of the proton
and neutron chemical potentials with pressure for both momentum
dependent and momentum independent Skyrme interactions.
A comparison of the thick curves (momentum dependent case) and
thin curves (momentum independent case) shows that the qualitative
behavior is the same. Quantitative difference exist with the momentum 
independent behavior being an enlargement of the momentum dependent shape.
The coexistence arc and the chemical instability loop meet at the cusp.
The behavior shown in these figures also confirm that
no truncation effects exist in our study.

\section{Summary and Conclusions}

  In this paper we studied the thermodynamic properties of a
two component system of hadronic matter made of protons and neutrons.
Our analysis is based on a mean field model using a local Skryme
interaction and includes both velocity or momentum dependent and
momentum independent interactions, besides volume, symmetry
and Coulomb effects. 
We have used a somewhat simplified description of the velocity
dependence of the nuclear interaction.
In particular we have used a density dependent effective mass approximation.
Effective mass approximations are frequently used in physics to
capture the main effects and they lead to a simpler set of equations
and a corresponding simpler analysis.
As noted we still keep Coulomb and surface terms which are present
in realistic nuclear systems.
It is the interplay of volume, surface, symmetry and Coulomb
and momentum dependent terms that is studied here.
In fact, the interplay of such terms makes nuclei a
unique system for studying phase transitions, chemical and
mechanical instability in binary systems.
We then applied the basic thermodynamic
relations that we developed to issues related to the
mechanical and chemical instability of nuclei and features
associated with a liquid/gas phase transition in this system.

Because of the two component nature of real nuclear systems,
the analysis involves a study of the behavior in proton fraction,
density and temperature ($y, \rho, T$)
and also proton fraction, pressure and temperature ($y, P, T$).
We studied systems with proton fraction $y = 0 \sim 1$, 
where $y = 0$ corresponds to a system of pure neutrons 
and $y = 1$ is for a system of pure protons.
An important system with large neutron excess is a neutron star.
The study of nuclear system with arbitrary proton/neutron ratios
is also important for future RIB experiments and for medium
energy collisions where the liquid/gas phase transition is
studied experimentally.
In a liquid/gas phase transition the liquid and gas phase have
different proton fractions because of symmetry and Coulomb effects.
The proton fraction in the liquid phase reflects a more symmetric
system than the gas phase where a higher asymmetry exists.
The process of producing a larger neutron excess in the gas phase
is referred to as isospin fractionation and a review can be
found in Ref.\cite{pr406,anp26,isosp,prl85xu,prl85li,nucl2877}. 
The process is modified somewhat by 
the Coulomb interaction which leads to proton diffusion of some 
protons from the liquid phase back into the gas phase as discussed 
in Ref.\cite{prc63,plb580}.

One of the unique aspects of the nuclear systems is a velocity
or momentum dependence in the two body interaction.
Here, we also study the role of this momentum dependence
first in the thermodynamic properties of the system.
Then, we extend the discussion of its role to nuclear
instabilities and phase transitions and make a comparison
with the case without momentum dependence.
A characteristic pattern of qualitative similarities and 
quantitative differences appear
between a momentum or velocity dependent Skryme interaction and a
momentum or velocity independent Skryme interaction. These patterns can
be seen in Fig.\ref{fig1}-\ref{fig9} and are discussed in detail
in Sect.\ref{rsltsec} which we briefly summarizes now. 

Fig.\ref{fig1} shows that the momentum dependence increases
the pressure at a given density. 
Fig.\ref{fig2} and Fig.\ref{fig4} show proton fraction
versus density and chemical potential versus density at several
pressures and at a fixed temperature. 
The qualitative features are the same between momentum
dependent and momentum independent forces. However, sizeable quantitative
differences are present between the two types of interactions. For
example the solid loops (momentum dependent interaction) in proton fraction 
versus density of Fig.\ref{fig2} are reduced versions of the same dashed
loops (momentum independent interaction). Similarly, the chemical
instability boundaries for a momentum dependent Skryme
interaction are found to be reduced versions of the same boundaries for
momentum independent Skyrme interactions as can be seen from Fig.\ref{fig5} 
and a comparison of the thin curves of Fig.\ref{fig6}--\ref{fig9} 
with the corresponding thick curves of these figures.
Fig.\ref{fig5} also shows that momentum dependent terms reduce
the height of the proton rich asymmetric loop ($y > y_E$)
and leave the height of the neutron asymmetric loop ($y < y_E$)
almost unchanged while the lowest pressure point,
which is the point of equal concentration $y_E$,
is shifted upward with the value of $y_E$ nearly unchanged.
From Fig.\ref{fig5} we also see that the chemical instability loop
lie on top of each other for proton and neutron as required by 
the general connection of Eq.(\ref{dmupn}).
Also seen is that the chemical instability loop is inside the coexistence
loop and tangent to it at the maxima of each loop.
The largest and smallest $y$ in the coexistence loops are
shifted inward towards the point of equal concentration $y_E$. 
Figs.\ref{fig6} and \ref{fig7} shows that the mechanical instability loop 
is inside the chemical instability loop
and tangent at the equal proton fraction $y_E(\rho)$ 
without touching it at the peak of them.

\acknowledgments
This work was supported in part by the US Department of Energy under DOE 
Grant No. DE-FG02-96ER-40987.
S.J.L. was on sabbatical leave from Kyung Hee University and spent 
a sabbatical year at Rutgers University in 2006-2007.

\end{document}